\documentclass[11pt]{article}
\usepackage{amsmath}   
\usepackage{amsfonts}    
\usepackage{mathrsfs} 
\usepackage{xfrac} 
\usepackage{tabu}
\usepackage{color}

\usepackage{tikz}
\usetikzlibrary{trees,snakes}
\usepackage{verbatim}

\usepackage{amsmath}
\usepackage{amsfonts}
\usepackage{amssymb}
\usepackage{graphicx}
\usepackage{tikz}
\usetikzlibrary{calc}
\usepackage{pgfplots}
\usepackage{cleveref}
\usepackage{graphicx}
\usepackage{subcaption}
\usetikzlibrary{trees,snakes}
\usepackage{verbatim}

\usepackage{mathrsfs} 
\usepackage{xfrac} 
\usepackage{tabu}
\usepackage{color}
\usepackage{tikz}
\usepackage{bbm}

\usepackage{amsmath}
\usepackage{amsfonts}
\usepackage{amssymb}
\usepackage{amsthm}
\usepackage{pgfplots}
\usepackage{cleveref}
\usepackage{graphicx}
\usepackage{marvosym}
\usepackage{mathrsfs} 
\usepackage{xfrac} 
\usepackage{tabu}
\usepackage{bbm}

\usepackage{tikz}
\usetikzlibrary{arrows,shapes,positioning}
\usetikzlibrary{calc}
\usepackage{pgfplots}
\usepackage{cleveref}
\usepackage{graphicx}
\usepackage{subcaption}
\usetikzlibrary{trees,snakes}
\usepackage{verbatim}
\usepackage{mathtools}
\usepackage{mathrsfs} 
\usepackage{xfrac} 
\usepackage{tabu}
\usepackage{color}
\usepackage{tikz}
\usepackage{mathtools}
\usepackage{tikz}
\usetikzlibrary{calc}
\usepackage{pgfplots}
\usepackage{cleveref}

\definecolor{myred}{RGB}{255, 0, 0}
\definecolor{myblue}{RGB}{0, 0, 255}
\newtheorem{theorem}{Theorem}
\newtheorem{lemma}{Lemma}

\newcommand{\code}{{\mathscr C}}

\newcommand{\eps}{\varepsilon}
\newcommand {\nn} {\nonumber}
\newcommand {\pr} {\mathbb{P}}
\newcommand{\IND}{\mathbbm{1}}

\newcommand{\DEF}{\overset{\Delta}{=}}

\newcommand{\dfn}{\stackrel{\triangle}{=}}
\newcommand {\exe} {\stackrel{\cdot} {=}}
\newcommand {\lexe} {\stackrel{\cdot} {\le}}
\newcommand {\gexe} {\stackrel{\cdot} {\ge}}

\newcommand {\bomega} {\mbox{\boldmath $\omega$}}
\newcommand {\reals} {{\rm I\!R}}

\newcommand {\hQ} {\hat{Q}}

\newcommand {\br} {\mbox{\boldmath $r$}}
\newcommand {\bs} {\boldsymbol{s}}

\newcommand {\bu} {\boldsymbol{u}}

\newcommand {\bv} {\mbox{\boldmath $v$}}
\newcommand {\bw} {\mbox{\boldmath $w$}}
\newcommand {\bx} {\boldsymbol{x}}
\newcommand {\by} {\boldsymbol{y}}
\newcommand {\bz} {\mbox{\boldmath $z$}}

\newcommand {\bS} {\boldsymbol{S}}

\newcommand {\bU} {\boldsymbol{U}}

\newcommand {\bX} {\boldsymbol{X}}

\newcommand {\bY} {\mbox{\boldmath $Y$}}
\newcommand {\bZ} {\mbox{\boldmath $Z$}}
\newcommand{\calA}{{\cal A}}
\newcommand{\calB}{{\cal B}}
\newcommand{\calC}{{\cal C}}

\newcommand{\calE}{{\cal E}}

\newcommand{\calG}{{\cal G}}

\newcommand{\calN}{{\cal N}}

\newcommand{\calP}{{\cal P}}
\newcommand{\calQ}{{\cal Q}}

\newcommand{\calS}{{\cal S}}
\newcommand{\calT}{{\cal T}}
\newcommand{\calU}{{\cal U}}

\newcommand{\calX}{{\cal X}}
\newcommand{\calY}{{\cal Y}}

\begin{document}
\thispagestyle{empty}
\title{Error Exponents of the Dirty-Paper and Gel'fand-Pinsker Channels\footnote{
		This research was supported by the Israel Science Foundation (ISF) grant no.\ 137/18.}\\}
\author{\\ Ran Tamir\footnote{R.\ Tamir is with the Department of Information Technology and Electrical Engineering, ETH Zurich, 8092 Z\"urich, Switzerland (e--mail: tamir@isi.ee.ethz.ch).} and Neri Merhav\footnote{N.\ Merhav is with the Andrew \& Erna Viterbi Faculty of Electrical and Computer Engineering, Technion -- Israel Institute of Technology, Technion City, Haifa 32000, Israel (e--mail: merhav@ee.technion.ac.il).}\\}
\maketitle
\vspace{1.5\baselineskip}
\setlength{\baselineskip}{1.5\baselineskip}

\begin{abstract}
We derive various error exponents for communication channels with random states, which are available non-causally at the encoder only.    
For both the finite-alphabet Gel'fand-Pinsker channel and its Gaussian counterpart, the dirty-paper channel, we derive random coding exponents, error exponents of the typical random codes (TRCs), and error exponents of expurgated codes. For the two channel models, we analyze some sub-optimal bin-index decoders, which turn out to be asymptotically optimal, at least for the random coding error exponent. 
For the dirty-paper channel, we show explicitly via a numerical example, that both the error exponent of the TRC and the expurgated exponent strictly improve upon the random coding exponent, at relatively low coding rates, which is a known fact for discrete memoryless channels without random states. 
We also show that at rates below capacity, the optimal values of the dirty-paper design parameter $\alpha$ in the random coding sense and in the TRC exponent sense are different from one another, and they are both different from the optimal $\alpha$ that is required for attaining the channel capacity.
For the Gel'fand-Pinsker channel, we allow for a variable-rate random binning code construction, and prove that the previously proposed maximum penalized mutual information decoder is asymptotically optimal within a given class of decoders, at least for the random coding error exponent. \\

\end{abstract}

\clearpage
\section{Introduction}

Channels with random states available as side information (SI) at the encoder have been studied for more than four decades. 
In 1980, Gel'fand and Pinsker \cite{GP1980} have derived the capacity formula for this channel model with discrete alphabets, and Costa has considered the Gaussian counterpart a few years later \cite{Costa1983}, which is widely known as the {\it dirty-paper channel} (DPC). 
Applications and extensions of their work include computer memories with defects \cite{HEG1983}, multiuser channel models \cite{Multiaccess, Broadcast}, joint source-channel coding \cite{MERHAV_Shamai_2003}, and information embedding in cover signals \cite{MoulinOsullivan2003}, just to name a few. Duality with source coding problems with SI was explored in \cite{Duality2, Duality1, Duality3}. 
For a tutorial on the subject, see \cite{KSM2007}.

Exponential error bounds have also been derived for channel models with random states as SI at the transmitter. Erez and Zamir \cite{ErezZamir2001} have derived random coding error exponents for the discrete memoryless additive noise channel. Later, Somekh--Baruch and Merhav \cite{SM2004} derived random coding error exponents for general discrete memoryless channels (DMCs) and for the multiple-access channel, both with channel SI at the transmitter. Error exponents for the DPC were studied by Lie, Moulin, and Koetter in \cite{LMK2006}. Random coding error exponents for an extended channel model with available (but not necessary identical) SI versions at the encoder, the adversary, and the decoder sides subject to various cost constrains were developed by Moulin and Wang \cite{Moulin2007}. A strong converse result and a sphere-packing type bound for the Gel'fand-Pinsker channel were derived by Tyagi and Narayan \cite{TN2009}.             

Error exponents of typical random codes (TRCs) and expurgated codes for the DPC and the Gel'fand-Pinsker channel are the main theme of this work. 
The error exponent of the TRC \cite{MERHAV_TYPICAL} is defined as\footnote{Note that this definition is different from the ordinary random-coding exponent, which is given by $E_{\mbox{\tiny r}}(R) = \lim_{n \to \infty} \left\{ - \tfrac{1}{n} \ln \mathbb{E} \left[P_{\mbox{\tiny e}}(\calC_{n}) \right] \right\}$.}  
\begin{align} \label{TRC_DEF}
E_{\mbox{\tiny trc}}(R) = \lim_{n \to \infty} \left\{- \tfrac{1}{n} \mathbb{E} \left[\ln P_{\mbox{\tiny e}}(\calC_{n}) \right] \right\},
\end{align}
where $R$ is the coding rate, $P_{\mbox{\tiny e}}(\calC_{n})$ is the error probability of a codebook $\calC_{n}$, and the expectation is with respect to (w.r.t.) the randomness of $\calC_{n}$ across the ensemble of codes. 
As explained in \cite{MERHAV_TYPICAL}, the error exponent of the TRC does not only strictly improve upon the random coding error exponent at low coding rates, but actually provides the exponent function around which the negative normalized logarithmic error probability concentrates.   

In \cite{BargForney}, Barg and Forney considered TRCs with independently and identically distributed codewords as well as typical linear codes, for the special case of the binary symmetric channel with maximum likelihood (ML) decoding. 
In \cite{PRAD2014} Nazari {\em et al.} provided bounds on the error exponents of TRCs for both DMCs (which coincide with one another for the optimal input distribution) and multiple--access channels. 
In a recent article by Merhav \cite{MERHAV_TYPICAL}, an exact single-letter expression has been derived for the error exponent of typical, random, fixed-composition codes, over DMCs, and a wide class of (stochastic) decoders, collectively referred to as the generalized likelihood decoder.
Later, Merhav has studied error exponents of TRCs for the colored Gaussian channel \cite{MERHAV_GAUSS}, typical random trellis codes \cite{MERHAV_TRELLIS}, and has derived a Lagrange--dual lower bound to the TRC exponent \cite{MERHAV_IID}. 
Recently, Tamir {\em et al.} have studied the large deviations behavior around the TRC exponent \cite{TMWG}, error exponents of typical random Slepian--Wolf codes \cite{TM}, 
and universal decoding for the TRC exponent \cite{TM_UD}.
More interesting concentration results of the logarithmic error probability of random codes around the error exponent of the TRC have been reported in \cite{TCFG2022}. 
Finally, a dual-domain lower bound to the TRC error exponent for general channels and pairwise-independent random-coding ensembles appears in \cite{CGF2022}.

For the DPC, an ordinary random binning code is analyzed and various error exponents are derived. Specifically, we provide some relatively simple expressions for the exact random coding error exponent and for the error exponent of the TRC. As in the ordinary DMC (without random states), an improvement is achieved at relatively low coding rates by code expurgation. 
We show also that in the DPC, the TRC exponent and the expurgated exponent strictly improve upon the random coding error exponent at some ranges of low rates. Although we implement a simple sub-optimal decoder, which seeks the closest codeword to the channel output vector, and not the optimal bin-index decoder, we show that at least for the random coding exponent, our decoder is as good as the optimal decoder. 
We also show that at low rates, such as rate zero, the optimal values of the design parameter\footnote{Recall that in the DPC, the transmitted vector is $\bx=\bu(m,\bs)-\alpha \bs$, with $\bs$ being the SI vector and $\bu(m,\bs)$ an auxiliary codeword that depends on both the SI vector and the transmitted message $m \in \{1,2,\ldots,M\}$, and where $\alpha$ is subject to optimization.} $\alpha$ in the random coding sense and in the TRC exponent sense are different, and they both differ from the optimal $\alpha$ that is required for achieving the channel capacity.     

Moving further, we turn to derive similar exponential error bounds for the Gel'fand-Pinsker channel. 
Also for the Gel'fand-Pinsker channel, we construct a random-binning code, but now we allow for a slightly more sophisticated code construction, which is in the spirit of \cite{Moulin2007}. 
For every possible SI type, we draw a different random-binning code, and select the binning rate individually, thus allowing for more degrees of freedom to improve the code performance. For completeness, we also adopt the penalized maximum mutual information (MMI) decoder proposed in \cite{Moulin2007}. 
The DPC and the Gel'fand-Pinsker channel are very close in spirit, and one may even consider the DPC as a special case of the Gel'fand-Pinsker channel, but still, they differ significantly by the simple fact that in the Gel'fand-Pinsker channel, we allow for a different sub-code for each SI type\footnote{Note that in the continuous case, the different types of $\bs$ are defined according to their norm, but $I(U;S)$ is independent of $\|\bs\|$. In the Gel'fand-Pinsker channel, on the other hand, there is no such parallel property.}.  
As a consequence, the analyses in the two cases follow different lines and the resulting expressions in the DPC case are considerably simpler, considering the low number of parameters that should undergo optimization. In the Gel'fand-Pinsker case, on the other hand, the final expressions are given by optimization problems, the dimension of which grows with the alphabet sizes. 
As opposed to our analysis for the DPC, which is tight at all coding rates, here, when deriving the TRC exponent and the expurgated bound, we make a pairwise error analysis, thus providing exponent functions which are tight only at relatively low rates, which is the range where these exponent functions improve upon the random coding error exponent (and the latter is tight at the high rates).
For the above described code construction of variable-rate binning (that depends on the empirical distribution of the SI), 
we prove that at least for the random coding error exponent, the penalized MMI decoder is actually optimal among all metrics that depend both on the joint empirical distribution of the codeword and the channel output sequence as well as on the SI possible type\footnote{Since the decoder has no access to the actual SI, it has to seek all codebooks pertaining to all possible SI types. Then, the penalized MMI decoder balances the fact that each codebook has a different binning rate.}. 
Due to recent findings in \cite{TM_UD}, we conjecture that the penalized MMI decoder is also optimal w.r.t.\ the error exponents of the TRC, but leave this question for future work.

The remaining part of the paper is organized as follows. 
In Section 2, we establish notation conventions. 
In Section 3, we consider the DPC; 
in Subsection 3.1, we formalize the model, review some preliminaries, and indicate the main objectives and in Subsection 3.2, we provide the main results and discuss them.    
Section 4 is devoted to the Gel'fand-Pinsker channel. 
In Subsection 4.1, we formalize the settings and in Subsection 4.2, we provide and discuss the results. 
In Section 5 we prove our results regarding the DPC and
in the Appendixes, we prove more supplementary results concerning DPC, discuss optimal bin index decoding, and prove the results concerning the Gel'fand-Pinsker model.

\section{Notation Conventions}

Throughout the paper, random variables will be denoted by capital letters and specific values they may take will be denoted by the corresponding lower case letters. Random vectors and their realizations will be denoted, respectively, by capital letters and the corresponding lower case letters, both in the bold face font. For example, the random vector $\bX = (X_1,X_2,\ldots,X_n)$, ($n$ -- positive integer) may take a specific vector value $\bx = (x_1,x_2,\ldots,x_n)$ in $\reals^{n}$. When used in the linear-algebraic context, these vectors should be thought of as column vectors, and so, when they appear
with superscript $T$, they will be transformed into row vectors by transposition. Thus, $\bx^{T}\by$ is
understood as the inner product of $\bx$ and $\by$. The notation $\|\bx\|$ will stand for the Euclidean norm of vector $\bx$. 
The $n$-dimensional hypersphere of radius $\sqrt{nr}$ will be denoted by $\calS(\sqrt{nr})$ and its surface area by $\mbox{Surf}\{\calS(\sqrt{nr})\}$.  
As customary in probability theory, we write $\bX=(X_{1}, \ldots, X_{n}) \sim \calN(0, \sigma^{2} \cdot I_{n})$ ($I_n$ being the $n\times n$ identity matrix) to denote that the probability density function of $\bX$ is 
\begin{align}
P_{\bX}(\bx) = (2\pi\sigma^{2})^{-n/2} \cdot \exp\left\{-\frac{1}{2\sigma^{2}} \|\bx\|^{2} \right\}.
\end{align} 
Logarithms, here and throughout the sequel, are taken to the natural base.
The probability of an event $\calE$ will be denoted by 
$\pr\{\calE\}$, and the expectation operator will be denoted by $\mathbb{E}[\cdot]$. 
For two positive sequences $a_{n}$ and $b_{n}$, the notation $a_{n} \doteq b_{n}$ will stand 
for equality in the exponential scale, that is, $\lim_{n \to \infty} \frac{1}{n} 
\ln \frac{a_{n}}{b_{n}} = 0$. 
Similarly, $a_{n} \lexe b_{n}$ means that $\limsup_{n \to \infty} (1/n) \ln \left(a_{n}/b_{n}\right) \leq 0$, and so on.
The indicator function of an event $\calE$ 
will be denoted by $\IND\{\calE\}$. 
The notation $[x]_{+}$ will stand for $\max \{0, x\}$. 

In the discrete case, alphabets of random variables and their realizations will be denoted by calligraphic letters. Accordingly, alphabets of random vectors and their realizations will be superscripted by their dimensions.  
For example, the random vector $\bX = (X_{1}, \dotsc , X_{n})$, may take a specific vector value $\bx = (x_{1}, \dotsc , x_{n})$ in $\calX^{n}$, 
the $n$-th order Cartesian power of $\mathcal{X}$, which is the alphabet of each component of this vector. Sources and channels will be subscripted by the names of the relevant random 
variables/vectors and their conditionings, whenever applicable, 
following the standard notation conventions, e.g., $Q_{X}$, $Q_{Y|X}$, and so on. 
When there is no room for ambiguity, these subscripts will be omitted. For a generic joint 
distribution $Q_{XY} = \{Q_{XY}(x,y), x \in \mathcal{X}, y \in \mathcal{Y} \}$, which will often 
be abbreviated by $Q$, information measures will be denoted in the conventional manner, but
with a subscript $Q$, that is, $H_{Q}(X)$ 
is the marginal entropy of $X$, $H_{Q}(X|Y)$ is the conditional entropy of $X$ given $Y$, 
$I_{Q}(X;Y) = H_{Q}(X) - H_{Q}(X|Y)$ is the mutual information between $X$ and $Y$, 
and similarly for other quantities. 
The weighted divergence between two conditional distributions (channels), say, $Q_{Y|X}$ and $W = \{W(y|x), 
x \in \calX, y \in \calY \}$, with weighting $Q_{X}$ is defined as
\begin{align}
D(Q_{Y|X} || W | Q_{X}) 
= \sum_{x \in \calX} Q_{X}(x) 
\sum_{y \in \calY} Q_{Y|X}(y|x) \ln \frac{Q_{Y|X}(y|x)}{W(y|x)}.
\end{align}

The empirical distribution of a sequence $\bx \in \mathcal{X}^{n}$, which will 
be denoted by $\hat{P}_{\bx}$, is the vector of relative frequencies, $\hat{P}_{\bx}(x)$, 
of each symbol $x \in \mathcal{X}$ in $\bx$. 
The type class of $\bx \in \mathcal{X}^{n}$, denoted $\calT^{n}(\bx)$, 
is the set of all vectors $\bx' \in \calX^{n}$ with $\hat{P}_{\bx'} = \hat{P}_{\bx}$. 
When we wish to emphasize the dependence of the type class on the empirical 
distribution $\hat{P}$, we will denote it by $\calT^{n}(\hat{P})$. 
The set of all types of vectors of length $n$ over $\calX$ will be denoted by $\calP_{n}(\calX)$.
Information measures 
associated with empirical distributions will be denoted with `hats' and will be subscripted 
by the sequences from which they are induced. 
Similar conventions will apply to the joint empirical distribution, 
the joint type class, the conditional empirical distributions and the conditional type classes 
associated with pairs (and multiples) of sequences of length $n$. 
Accordingly, $\hat{P}_{\bx\by}$ would be the joint empirical distribution 
of $(\bx, \by) = \{(x_{i}, y_{i})\}_{i=1}^{n}$, $\calT^{n}(Q_{X|Y}|\by)$ will stand for the 
conditional type class induced by a sequence $\by$ and a relevant empirical conditional distribution $Q_{X|Y}$, which is the set of all vectors $\bx \in \calX^{n}$ with $\hat{P}_{\bx\by} = Q_{X|Y} \times \hat{P}_{\by}$, $\hat{I}_{\bx\by}(X;Y)$ will denote the empirical mutual information induced by $\bx$ and $\by$, and so on. 
Similar conventions will apply to triplets of sequences, say, $\{(\bx, \by, \bz)\}$, etc. 
Likewise, when we wish to emphasize the dependence of empirical information measures upon a given empirical distribution given by $Q$, we denote them using the subscript $Q$, as described above.

\section{Error Exponents of Dirty--Paper Coding}

\subsection{Setup, Preliminaries, and Objectives}
\label{SEC_Objectives}
Consider the DPC,
\begin{equation}
\bY=\bX+\bS+\bZ,
\end{equation}
where $\bX=(X_1,\ldots,X_n)$ is the channel input vector, whose power is limited according to 
$\|\bX\|^2\le nP$,
$\bS=(S_1,\ldots,S_n)\sim\calN(0,Q\cdot I_n)$, is the random interference signal vector, and
$\bZ=(Z_1,\ldots,Z_n)\sim\calN(0,\sigma_Z^2\cdot I_n)$ is the additive Gaussian noise, which is independent of
$\bX$ and $\bS$. Here, $Q>0$ and $\sigma_Z^2>0$ are the interference 
variance and the noise variance, respectively.

Consider the following mechanism of random codebook selection. Let the coding rate, 
$R\ge 0$, be given. 
Let $\alpha > 0$ be a design parameter (to be optimized later\footnote{Note that the optimal $\alpha$ here may not be the same as the optimal $\alpha$ that achieves the capacity \cite{Costa1983}.}), and define
\begin{equation} \label{DEF_rho0}
\rho_0\dfn\sqrt{\frac{\alpha^2Q}{P+\alpha^2Q}}=\sqrt{\frac{\alpha^2Q}{W}},
\end{equation}
where $W\dfn P+\alpha^2Q$. Next, define
\begin{equation}
I_{US}\dfn\frac{1}{2}\ln\left(\frac{1}{1-\rho_0^2}\right)=\frac{1}{2}\ln\left(\frac{W}{P}\right).
\end{equation}
Generate a random codebook, $\calC$, 
of $M=e^{n(R+I_{US}+\Delta)}$ ($\Delta > 0$ being arbitrarily small) codewords, 
$\bu_k$, $k=0,1,\ldots,M-1$, by independent random selection of each codeword
under the uniform distribution over the surface of an $n$-dimensional hypsphere of radius 
$\sqrt{nW}$, centered at the origin. Next, divide
$\calC$ into $M_1=e^{nR}$ bins, $\calC_0,\calC_1,\ldots,C_{M_1-1}$, 
each of size $M_2=M/M_1=e^{n(I_{US}+\Delta)}$.

Let $\epsilon > 0$ be arbitrarily small. Given a message $m\in\{0,1,\ldots,M_1-1\}$ and a realization, 
$\bs$ of $\bS$, the encoder seeks, within bin number $m$, a codeword,
$\bu_k$, $k\in\{M_2m,M_2m+1,\ldots, M_2(m+1)-1\}$, such that 
\begin{equation}
\label{ustype}
\alpha_{\bs}(1+\epsilon)\|\bs\|^2\ge\bu_k^T\bs\ge\alpha_{\bs}\|\bs\|^2.
\end{equation} 
The coefficient $\alpha_{\bs}$ in \eqref{ustype} is defined as
\begin{equation}
\alpha_{\bs}\dfn\alpha\cdot\sqrt{\frac{nQ}{\|\bs\|^2}}.
\end{equation}
We denote
\begin{equation}
\calT(\bU|\bs)=\{\bu:~\|\bu\|^2=nW,~\alpha_{\bs}(1+\epsilon)\ge\bu^T\bs\ge\alpha_{\bs}\|\bs\|^2\}.
\end{equation}
The criterion \eqref{ustype} is equivalent to seeking a codeword $\bu_k$ 
whose empirical correlation coefficient with $\bs$, 
$\hat{\rho}(\bu_k,\bs)\dfn\bu_k^T\bs/[\|\bu_k\|\cdot\|\bs\|]$, is nearly $\rho_0$.
Since the bin size, $M_2$, is exponentially larger than $e^{nI_{US}}$, 
the probability of not finding even one codeword in bin
number $m$, which satisfies this condition, decays double exponentially with $n$. 
We therefore safely neglect this event of encoding failure.
We define the encoder,
more precisely, as follows: suppose that for the given $\bs$ and $m$ observed by the encoder, there are
$K_n(\bs,m)\ge 1$ codewords within the bin $\calC_m$ that satisfy \eqref{ustype}.
Then, the encoder
randomly selects one of them under the uniform distribution, 
to be the transmitted codeword, $\bu_k$, i.e.,
\begin{equation}
\label{se}
\pr\{\bu_k=\bu|\bs,m\}=\frac{1}{K_n(\bs,m)},~~~\forall~
\bu\in\calC_m\cap\calT(\bU|\bs).
\end{equation}
Upon randomly selecting $\bu_k$, the encoder
transmits\footnote{One may also consider a more general encoding scheme, where $\bx$ is randomly selected from the conditional type class
\begin{align}
\calT(\bx|\bu,\bs)=\left\{\bx:~\|\bx\|^2=nP,~\bx^T\bu=n\sqrt{PW}\rho_{xu},~\bx^T\bs=
n\sqrt{P}\hat{\sigma}_s\rho_{xs}\right\},
\end{align}
where the parameters $\rho_{xs}$ and $\rho_{xu}$ are subjected to optimization.
Note that this more general coding is much closer in spirit to the coding strategy for the Gel'fand-Pinsker channel, as given in Subsection \ref{SEC_GP_setup}, where the transmitted codeword is drawn from the conditional type class given $\bu,\bs$.} 
\begin{equation} \label{DPC_encoder}
\bx=\bu_k-\alpha_{\bs}\cdot\bs. 
\end{equation}
Note that
\begin{eqnarray}
\|\bx\|^2&=&\|\bu_k\|^2-2\alpha_{\bs}\bu_k^T\bs+\alpha_{\bs}^2\|\bs\|^2\nonumber\\
&\le&nW-2\alpha_{\bs}^2\|\bs\|^2+\alpha_{\bs}^2\|\bs\|^2\nonumber\\
&=&nW-\alpha_{\bs}^2\|\bs\|^2\nonumber\\
&=&nW-n\alpha^2Q\nonumber\\
&=&nP,
\end{eqnarray}
and so, the power constraint is met.

The receiver implements the MMI decoder, which in the
Gaussian-quadratic case considered here, amounts to seeking the codeword, 
$\bu_k$, $k\in\{0,1,\ldots, M-1\}$, that maximizes
the squared (or, equivalently, the absolute) empirical correlation, $\hat{\rho}^2(\bu_k,\by)$, i.e.,
\begin{equation}
\label{decodek}
\hat{k}=\mbox{arg}\max_k\hat{\rho}^2(\bu_k,\by)=
\mbox{arg}\max_k\frac{(\bu_k^T\by)^2}{\|\bu_k\|^2\cdot\|\by\|^2}
=\mbox{arg}\max_k(\bu_k^T\by)^2,
\end{equation}
and finally, decoding $m$ according to the bin to which $\bu_{\hat{k}}$ belongs, that is,
\begin{equation}
\label{decodem}
\hat{m}=\left\lfloor \frac{\hat{k}}{M_2}\right\rfloor.
\end{equation}
This two--stage decoding process is asymptotically optimal, at least in the random-coding sense, as it follows by a combination of two arguments. First, we argue by the end of Subsection \ref{Sec_Preparatory_Steps} ahead that MMI decoding is asymptotically optimal, provided that one wants to fully decode $\bu_k$ (and not only the bin), and second, we prove in Appendix A that a simple ML decoder and the optimal bin index decoder are asymptotically equivalent, at least in the random coding sense.   

For a given code, let $P_{\mbox{\tiny e}}(\calC_{n})$ be the probability of error in the bin index decoding.  
The random coding error exponent is defined as
\begin{align} \label{DEF_RCE_LOWER}
\mathsf{E}_{\mbox{\tiny r}}(R)
=\lim_{n \to \infty} -\frac{1}{n} \ln \mathbb{E} \left[ P_{\mbox{\tiny e}}(\calC_{n}) \right],  
\end{align}
and the error exponent of the TRC is defined by
\begin{align} \label{DEF_TRC}
\mathsf{E}_{\mbox{\tiny trc}}(R)
=\lim_{n \to \infty} -\frac{1}{n} \mathbb{E} \left[\ln P_{\mbox{\tiny e}}(\calC_{n}) \right].  
\end{align}
Our first objective is to derive exact single-letter expressions for \eqref{DEF_RCE_LOWER} and \eqref{DEF_TRC}.

Another objective is to prove the existence of a sequence of codes 
$\code = \{\calC_{n}\}_{n=1}^{\infty}$, whose error exponent is strictly higher than $\mathsf{E}_{\mbox{\tiny r}}(R)$ and $\mathsf{E}_{\mbox{\tiny trc}}(R)$, at least at low coding rates, and obtain a single--letter expression that lower bounds the following limit
\begin{align}
\mathsf{E}_{\mbox{\tiny ex}}(\code) 
= \liminf_{n \to \infty} - \frac{1}{n} \ln \max_{m} P_{\mbox{\tiny e}|m}(\calC_{n}), 
\end{align}
where $P_{\mbox{\tiny e}|m}(\calC_{n})$ is the conditional error probability, given that message $m$ was transmitted. 

In this section, we analyze, study and discuss 
the random coding exponent, the typical code exponent and the expurgated 
exponent associated with this model, and we examine them, not only as functions of the rate $R$, but also as
functions of $\alpha$ and of $Q$. In particular, we show that the optimal value of $\alpha$ for rates below
capaity may be different from the optimal $\alpha$ that is needed to achieve capacity, namely,
$\alpha^*=P/(P+\sigma_Z^2)$ \cite[eq.\ (7)]{Costa1983}. 

For the sake of convenience and simplicity, we initially assume that the random interference vector,
$\bS$, is distributed uniformly across the surface of the $n$-dimensional hypersphere of radius $\sqrt{n\hat{Q}}$
(that is, $\|\bS\|^2=n\hat{Q}$ with probability one) and return to the Gaussian model later (by weighing the various
radii according to the Gaussian divergence between $\hat{Q}$ and $Q$), as part of our main theorem
in this section. Recall that the encoder depends on $\bS$ only via its empirical correlations with the various
codewords, $\{\bu_k\}$, and these empirical correlations are scale--invariant. The decoder, on the other hand, has
no access to $\bS$ anyway.

\subsection{Statement of the Main Results} \label{Sec_DPC_Results}

Let $P$, $Q$, $\hat{Q}$, $\sigma^2$ and $\alpha\in[0,1]$ be given.
Define the following quantities:
\begin{eqnarray}
z&=&1-e^{-2(R+I_{US})}=1-\frac{Pe^{-2R}}{W}\\
a&=&\alpha\sqrt{Q}(\sqrt{\hat{Q}}-\alpha\sqrt{Q})+W,
\end{eqnarray}
and for two given reals, $\rho$ and $\varrho$, both in $(-1,1)$, define also
\begin{equation}
b=\sqrt{W}(\sqrt{\hat{Q}}-\alpha\sqrt{Q})\varrho+\rho W.\\
\end{equation}
For two given positive reals, $p$ and $q$ ($p\le q$), define
\begin{equation}
\mu(p,q)=1+(q-p)z
\end{equation}
\begin{equation}
\Delta(p,q)=[\mu(p,q)+p]\cdot[\mu(p,q)-q]+\rho^2pq.
\end{equation}
Next, define
\begin{eqnarray}
E(R,\hat{Q},\alpha,\rho,\varrho,p,q)&=&
\frac{1}{2}\ln\mu(p,q)+\frac{1}{2\sigma^2\mu(p,q)}\bigg[
(q-p)z[W+\hat{Q}-\alpha^2Q]+\nonumber\\
& &\frac{\mu(p,q)(pa^2-qb^2)-pq(a^2-2\rho ab+b^2)}
{W\cdot\Delta(p,q)}\bigg]\\
\label{Inner_Optimization}
E(R,\hat{Q},\alpha,\rho,\varrho)&=&\sup_{\{(p,q):~0\le p\le q,~\Delta(p,q)> 0\}}
E(R,\hat{Q},\alpha,\rho,\varrho,p,q).
\end{eqnarray}
Define also $\calP(\rho,\rho_0)$ as the set of values
of $\varrho$ within $(-1,1)$ such that the correlation matrix
\begin{align}
\left(\begin{array}{ccc}
1 & \rho & \rho_0\\
\rho & 1 & \varrho\\
\rho_0 & \varrho & 1\end{array}\right)
\end{align}
would be positive semi-definite, and let
\begin{equation}
E(R,\hat{Q},\alpha,\rho)=\inf_{\varrho\in\calP(\rho,\rho_0)}
\left\{E(R,\hat{Q},\alpha,\rho,\varrho)+L(\rho,\varrho)\right\}+\frac{1}{2}\ln(1-\rho^2)
-\frac{1}{2}\ln\frac{W}{P},
\end{equation}
where
\begin{equation}
L(\rho,\varrho)=-\frac{1}{2}\ln(1-\rho^2-\varrho^2-\rho_0^2+2\rho\varrho\rho_0).
\end{equation}
Finally, letting
\begin{eqnarray}
E_0(R,\hat{Q},\alpha,\rho)&\dfn&E(R,\hat{Q},\alpha,\rho)+
\frac{1}{2}\ln\frac{1}{1-\rho^2}-\frac{1}{2}\ln\frac{W}{P}-R\nonumber\\
&=&\inf_{\varrho\in\calP(\rho,\rho_0)}\left\{E(R,\hat{Q},\alpha,\rho,\varrho)+L(\rho,\varrho)\right\}
-\ln\frac{W}{P}-R,
\end{eqnarray}
we have the following result, the proof of which is given in Section \ref{Sec_Proof_DPC}.
\begin{theorem} \label{Theorem_DPC}
	Consider the setting described in Subsection \ref{SEC_Objectives}.
	The error exponents $\mathsf{E}_{\mbox{\tiny r}}(R)$ and $\mathsf{E}_{\mbox{\tiny trc}}(R)$ associated with dirty-paper coding
	and MMI decoding are given by: 
	\begin{eqnarray}
	E_{\mbox{\tiny r}}(R,\hat{Q},\alpha)&=&\inf_{\rho\in(-1,1)}E_0(R,\hat{Q},\alpha,\rho), \\
	E_{\mbox{\tiny trc}}(R,\hat{Q},\alpha)&=&\inf_{\rho^2<1-(P/W)^2e^{-4R}}E_0(R,\hat{Q},\alpha,\rho),
	\end{eqnarray}
	respectively.
	In addition, there exists a sequence of codes, $\code=\{\calC_{n}\}_{n=1}^{\infty}$, such that
	the error exponent $\mathsf{E}_{\mbox{\tiny ex}}(\code)$
	associated with dirty-paper coding
	and MMI decoding is lower-bounded by:
	\begin{eqnarray}
	E_{\mbox{\tiny ex}}(R,\hat{Q},\alpha)&=&\inf_{\rho^2<1-(P/W)^2e^{-2R}}E_0(R,\hat{Q},\alpha,\rho).
	\end{eqnarray}
	If $\bS$ is distributed uniformly over the surface of a hypersphere of radius $\sqrt{nQ}$, take $\hat{Q}=Q$ in all above
	expressions, and then maximize over the design parameter $\alpha\in[0,1]$ 
	for each error exponent, to obtain optimal performance.
	Alternatively, if $\bS\sim\calN(0,Q\cdot I_n)$, first minimize over $\hQ$ 
	each one of the corresponding error exponents plus the	additional term 
	\begin{equation}
	D(\hat{Q}\|Q)=\frac{1}{2}\left[\frac{\hat{Q}}{Q}-\ln\left(\frac{\hat{Q}}{Q}\right)-1\right],
	\end{equation}
	and finally, maximize again over $\alpha$.
\end{theorem}

We can simplify the bounds significantly by confining the inner--most optimization \eqref{Inner_Optimization} to $p=q$,
corresponding to a relatively simple union--bound analysis of pairwise error probabilities, which is tight at some 
range of low rates. This is relevant since the interesting range where the TRC error exponent differs from the
random coding error exponent is the range of low rates anyway.
Accordingly, the choice $p=q$ yields:
\begin{equation}
\label{Simp1}
E(R,\hat{Q},\alpha,\rho,\varrho,q,q)=
\frac{q(a^2-b^2)-q^2(a^2-2\rho ab+b^2)}
{2W\sigma^2[1-q^2(1-\rho^2)]}
\end{equation}
to be maximized over $q\in[0,1/\sqrt{1-\rho^2})$
which can be carried out explicitly, yielding
\begin{equation}
E(R,\hat{Q},\alpha,\rho,\varrho)\ge\frac{1}{2W\sigma^2\sqrt{1-\rho^2}}\cdot 
T\left(a^2-b^2,\frac{a^2-2\rho ab+b^2}
{\sqrt{1-\rho^2}}\right),
\end{equation}
where the function $T(\cdot,\cdot)$ is defined as
\begin{equation}
\label{Simp2}
T(g,h)\dfn\sup_{0\le\tau<1}\frac{g\tau-h\tau^2}{1-\tau^2}=\left\{\begin{array}{ll}
0 & g\le 0\\
\frac{g^2}{2(h+\sqrt{h^2-g^2})} & h\ge g> 0\\
\infty & g > h\ge 0\end{array}\right. .
\end{equation}

\begin{figure}[ht!]
	\centering
	\begin{tikzpicture}[scale=1.2]
	\begin{axis}[
	disabledatascaling,
	scaled x ticks=false,
	xticklabel style={/pgf/number format/fixed,
		/pgf/number format/precision=3},
	scaled y ticks=false,
	yticklabel style={/pgf/number format/fixed,
		/pgf/number format/precision=3},
	xlabel={$R$},
	ylabel={Error exponents},
	xmin=0, xmax=0.6,
	ymin=0.4, ymax=2,
	xtick={0,0.1,0.2,0.3,0.4,0.5},
	legend pos=north east,
	]
	
	\addplot[smooth,color=black!30!green,thick]
	table[row sep=crcr] 
	{
0	1.01E+00	\\
0.01	9.98E-01	\\
0.02	9.88E-01	\\
0.03	9.78E-01	\\
0.04	9.68E-01	\\
0.05	9.58E-01	\\
0.06	9.48E-01	\\
0.07	9.38E-01	\\
0.08	9.28E-01	\\
0.09	9.18E-01	\\
0.1	9.08E-01	\\
0.11	8.98E-01	\\
0.12	8.88E-01	\\
0.13	8.78E-01	\\
0.14	8.68E-01	\\
0.15	8.58E-01	\\
0.16	8.48E-01	\\
0.17	8.38E-01	\\
0.18	8.28E-01	\\
0.19	0.818109137	\\
0.2	0.808109137	\\
0.21	0.798109137	\\
0.22	0.788109137	\\
0.23	7.78E-01	\\
0.24	0.768109137	\\
0.25	0.758109137	\\
0.26	0.748109137	\\
0.27	0.738109137	\\
0.28	0.728109137	\\
0.29	0.718109137	\\
0.3	0.708109137	\\
0.31	6.98E-01	\\
0.32	6.88E-01	\\
0.33	6.78E-01	\\
0.34	6.68E-01	\\
0.35	6.58E-01	\\
0.36	6.48E-01	\\
0.37	6.38E-01	\\
0.38	6.28E-01	\\
0.39	6.18E-01	\\
0.4	6.08E-01	\\
0.41	5.98E-01	\\
0.42	5.88E-01	\\
0.43	5.78E-01	\\
0.44	5.68E-01	\\
0.45	5.58E-01	\\
0.46	5.48E-01	\\
0.47	5.38E-01	\\
0.48	0.528109137	\\
0.49	5.18E-01	\\
0.5	5.08E-01	\\
0.51	0.498109137	\\
0.52	0.488109137	\\
0.53	0.478109137	\\
0.54	0.468109137	\\
0.55	0.458109137	\\
0.56	0.448109137	\\
0.57	0.438109137	\\
0.58	0.428109137	\\
0.59	0.418109137	\\
0.6	0.408109137	\\
	};
	\legend{}
	\addlegendentry{$E_{\mbox{\tiny r}}(R)$}

	\addplot[smooth,color=black!20!orange,thick,dash pattern={on 2pt off 1pt}]
	table[row sep=crcr] 
	{
0	1.89E+00	\\
0.01	1.72E+00	\\
0.02	1.60E+00	\\
0.03	1.51E+00	\\
0.04	1.43E+00	\\
0.05	1.36E+00	\\
0.06	1.30E+00	\\
0.07	1.24E+00	\\
0.08	1.19E+00	\\
0.09	1.14E+00	\\
0.1	1.10E+00	\\
0.11	1.06E+00	\\
0.12	1.03E+00	\\
0.13	9.98E-01	\\
0.14	9.68E-01	\\
0.15	9.41E-01	\\
0.16	9.16E-01	\\
0.17	8.92E-01	\\
0.18	8.71E-01	\\
0.19	8.51E-01	\\
0.2	8.32E-01	\\
0.21	8.15E-01	\\
0.22	8.00E-01	\\
0.23	7.85E-01	\\
0.24	7.72E-01	\\
0.25	7.60E-01	\\
0.26	7.48E-01	\\
0.27	7.38E-01	\\
0.28	7.28E-01	\\
0.29	0.718124189	\\
0.3	0.708066243	\\
0.31	6.98E-01	\\
0.32	6.88E-01	\\
0.33	6.78E-01	\\
0.34	6.68E-01	\\
0.35	6.58E-01	\\
0.36	6.48E-01	\\
0.37	6.38E-01	\\
0.38	6.28E-01	\\
0.39	6.18E-01	\\
0.4	6.08E-01	\\
0.41	5.98E-01	\\
0.42	5.88E-01	\\
0.43	5.78E-01	\\
0.44	5.68E-01	\\
0.45	5.58E-01	\\
0.46	5.48E-01	\\
0.47	5.38E-01	\\
0.48	5.28E-01	\\
0.49	5.18E-01	\\
0.5	5.08E-01	\\
0.51	4.98E-01	\\
0.52	4.88E-01	\\
0.53	4.78E-01	\\
0.54	4.68E-01	\\
0.55	4.58E-01	\\
0.56	4.48E-01	\\
0.57	4.38E-01	\\
0.58	0.42801674	\\
0.59	0.418019057	\\
0.6	0.408059924	\\
	};
	\addlegendentry{$E_{\mbox{\tiny trc}}(R)$}

	\addplot[smooth,color=black!20!purple,thick,dash pattern={on 3pt off 2pt}]
	table[row sep=crcr] 
	{
0	1.89E+00	\\
0.01	1.78E+00	\\
0.02	1.71E+00	\\
0.03	1.64E+00	\\
0.04	1.58E+00	\\
0.05	1.53E+00	\\
0.06	1.48E+00	\\
0.07	1.43E+00	\\
0.08	1.39E+00	\\
0.09	1.35E+00	\\
0.1	1.31E+00	\\
0.11	1.27E+00	\\
0.12	1.24E+00	\\
0.13	1.20E+00	\\
0.14	1.17E+00	\\
0.15	1.14E+00	\\
0.16	1.11E+00	\\
0.17	1.08E+00	\\
0.18	1.05E+00	\\
0.19	1.03E+00	\\
0.2	1.00E+00	\\
0.21	9.78E-01	\\
0.22	9.55E-01	\\
0.23	9.32E-01	\\
0.24	9.10E-01	\\
0.25	8.88E-01	\\
0.26	8.68E-01	\\
0.27	8.48E-01	\\
0.28	8.28E-01	\\
0.29	8.09E-01	\\
0.3	7.91E-01	\\
0.31	7.73E-01	\\
0.32	7.56E-01	\\
0.33	7.39E-01	\\
0.34	7.22E-01	\\
0.35	7.06E-01	\\
0.36	6.91E-01	\\
0.37	6.76E-01	\\
0.38	6.61E-01	\\
0.39	6.46E-01	\\
0.4	6.32E-01	\\
0.41	6.19E-01	\\
0.42	6.05E-01	\\
0.43	5.92E-01	\\
0.44	5.80E-01	\\
0.45	5.67E-01	\\
0.46	5.55E-01	\\
0.47	5.43E-01	\\
0.48	0.531962368	\\
0.49	5.21E-01	\\
0.5	0.509674536	\\
0.51	0.49889998	\\
0.52	0.48838102	\\
0.53	0.478088564	\\
0.54	0.468015507	\\
0.55	0.458155001	\\
0.56	0.44802727	\\
0.57	0.438046365	\\
0.58	0.428124189	\\
0.59	0.418015196	\\
0.6	0.408066243	\\
	};
	\addlegendentry{$E_{\mbox{\tiny ex}}(R)$}	
	
	\end{axis}
	
	\end{tikzpicture}
	\caption{Low--rate error exponent functions for the DPC with the parameters, $P=10$, $Q=\sigma_Z^2=1$, and with optimized $\alpha$ in the interval $[0,1]$.} \label{graph1}
\end{figure}
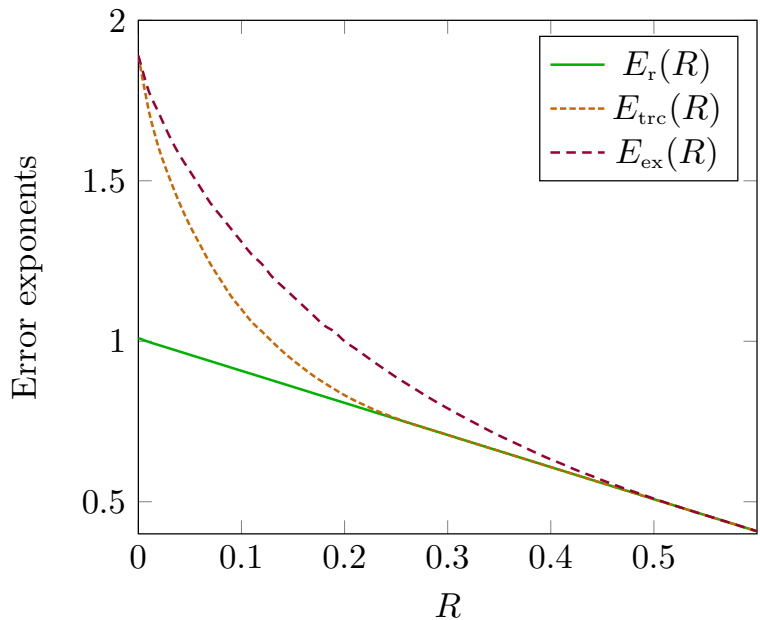

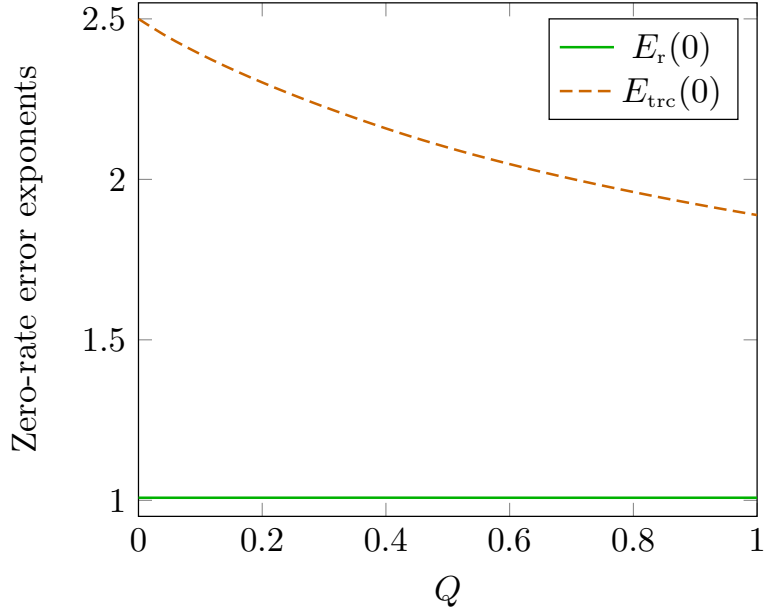
\begin{figure}[ht!]
	\centering
	\begin{tikzpicture}[scale=1.2]
	\begin{axis}[
	disabledatascaling,
	scaled x ticks=false,
	xticklabel style={/pgf/number format/fixed,
		/pgf/number format/precision=3},
	scaled y ticks=false,
	yticklabel style={/pgf/number format/fixed,
		/pgf/number format/precision=3},
	xlabel={$Q$},
	ylabel={Zero-rate error exponents},
	xmin=0, xmax=1,
	ymin=0.95, ymax=2.55,
	legend pos=north east,
	]
	
	\addplot[smooth,color=black!30!green,thick]
	table[row sep=crcr] 
	{
0	1.007980964	\\
0.04	1.0080004	\\
0.08	1.007988157	\\
0.12	1.008005756	\\
0.16	1.008034022	\\
0.2	1.008066152	\\
0.24	1.008071946	\\
0.28	1.008097711	\\
0.32	1.008135108	\\
0.36	1.008183241	\\
0.4	1.008164274	\\
0.44	1.008143579	\\
0.48	1.008107173	\\
0.52	1.008056483	\\
0.56	1.00803151	\\
0.6	1.008041123	\\
0.64	1.00800555	\\
0.68	1.007987753	\\
0.72	1.008021347	\\
0.76	1.007983614	\\
0.8	1.008001262	\\
0.84	1.008008978	\\
0.88	1.008013525	\\
0.92	1.008052669	\\
0.96	1.008052338	\\
1	1.008109137	\\
	};
	\legend{}
	\addlegendentry{$E_{\mbox{\tiny r}}(0)$}

	\addplot[smooth,color=black!20!orange,thick,dash pattern={on 4pt off 2pt}]
	table[row sep=crcr] 
	{
0	2.5	\\
0.04	2.451914614	\\
0.08	2.409739294	\\
0.12	2.371222415	\\
0.16	2.335487696	\\
0.2	2.302032951	\\
0.24	2.27047909	\\
0.28	2.240558945	\\
0.32	2.212064291	\\
0.36	2.184798639	\\
0.4	2.158661892	\\
0.44	2.133813432	\\
0.48	2.110437261	\\
0.52	2.088381198	\\
0.56	2.067475181	\\
0.6	2.047608301	\\
0.64	2.028669788	\\
0.68	2.010563201	\\
0.72	1.993231047	\\
0.76	1.976596137	\\
0.8	1.960601636	\\
0.84	1.945196029	\\
0.88	1.930354621	\\
0.92	1.916015063	\\
0.96	1.902147472	\\
1	1.888731132	\\
	};
	\addlegendentry{$E_{\mbox{\tiny trc}}(0)$}	
	
	\end{axis}
	
	\end{tikzpicture}
	\caption{Zero--rate error exponents as functions of $Q$ for the DPC with the parameters, $P=10$, $\sigma_Z^2=1$, and 
	with optimized $\alpha$ in the interval $[0,1]$. The random coding exponent, $E_{\mbox{\tiny r}}(0)$ and the typical--code error exponent, $E_{\mbox{\tiny trc}}(0)$, are in solid and dashed lines, respectively.		
	The exponents $E_{\mbox{\tiny trc}}(0)$ and $E_{\mbox{\tiny ex}}(0)$ coincide.} \label{graph2}
\end{figure}

\begin{figure}[ht!]
	\centering
	\begin{tikzpicture}[scale=1.2]
	\begin{axis}[
	disabledatascaling,
	scaled x ticks=false,
	xticklabel style={/pgf/number format/fixed,
		/pgf/number format/precision=3},
	scaled y ticks=false,
	yticklabel style={/pgf/number format/fixed,
		/pgf/number format/precision=3},
	xlabel={$\alpha$},
	ylabel={Zero-rate error exponents},
	xmin=0, xmax=1,
	ymin=0.6, ymax=2,
	legend pos=south east,
	]
	
	\addplot[smooth,color=black!30!green,thick]
	table[row sep=crcr] 
	{
0.01	7.71E-01	\\
0.02	7.75E-01	\\
0.03	7.79E-01	\\
0.04	7.83E-01	\\
0.05	7.87E-01	\\
0.06	7.91E-01	\\
0.07	7.95E-01	\\
0.08	7.99E-01	\\
0.09	8.03E-01	\\
0.1	8.06E-01	\\
0.11	8.10E-01	\\
0.12	8.14E-01	\\
0.13	8.18E-01	\\
0.14	8.22E-01	\\
0.15	8.26E-01	\\
0.16	8.29E-01	\\
0.17	8.33E-01	\\
0.18	8.37E-01	\\
0.19	0.840858429	\\
0.2	0.844544288	\\
0.21	0.848194276	\\
0.22	0.851862701	\\
0.23	8.56E-01	\\
0.24	0.859245282	\\
0.25	0.862954736	\\
0.26	0.866639225	\\
0.27	0.870148697	\\
0.28	0.873664925	\\
0.29	0.87718864	\\
0.3	0.880717545	\\
0.31	8.84E-01	\\
0.32	8.88E-01	\\
0.33	8.91E-01	\\
0.34	8.95E-01	\\
0.35	8.98E-01	\\
0.36	9.01E-01	\\
0.37	9.05E-01	\\
0.38	9.08E-01	\\
0.39	9.11E-01	\\
0.4	9.15E-01	\\
0.41	9.18E-01	\\
0.42	9.21E-01	\\
0.43	9.24E-01	\\
0.44	9.27E-01	\\
0.45	9.30E-01	\\
0.46	9.33E-01	\\
0.47	9.36E-01	\\
0.48	0.939145368	\\
0.49	9.42E-01	\\
0.5	9.45E-01	\\
0.51	0.947651449	\\
0.52	0.950362936	\\
0.53	0.953042869	\\
0.54	0.955689561	\\
0.55	0.958301365	\\
0.56	0.960876673	\\
0.57	0.963413921	\\
0.58	0.965911581	\\
0.59	0.968368168	\\
0.6	0.970714666	\\
0.61	0.97293789	\\
0.62	0.975115689	\\
0.63	0.977246701	\\
0.64	0.979329609	\\
0.65	0.981363133	\\
0.66	0.983346035	\\
0.67	0.985277115	\\
0.68	0.987155215	\\
0.69	0.98894009	\\
0.7	0.990661481	\\
0.71	0.992324334	\\
0.72	0.993927708	\\
0.73	0.995438883	\\
0.74	0.996782847	\\
0.75	0.998059997	\\
0.76	0.999274023	\\
0.77	1.000424173	\\
0.78	1.001509731	\\
0.79	1.00253002	\\
0.8	1.003468143	\\
0.81	1.004288015	\\
0.82	1.005039298	\\
0.83	1.005721524	\\
0.84	1.006334262	\\
0.85	1.006877112	\\
0.86	1.007274353	\\
0.87	1.007598809	\\
0.88	1.00785136	\\
0.89	1.008031803	\\
0.9	1.008109137	\\
0.91	1.008062706	\\
0.92	1.007943067	\\
0.93	1.007750206	\\
0.94	1.007475757	\\
0.95	1.007051665	\\
0.96	1.006554132	\\
0.97	1.005983322	\\
0.98	1.005339425	\\
0.99	1.00454051	\\
1	1.003665462	\\
	};
	\legend{}
	\addlegendentry{$E_{\mbox{\tiny r}}(0)$}

	\addplot[smooth,color=black!20!orange,thick,dash pattern={on 4pt off 2pt}]
	table[row sep=crcr] 
	{
0.01	1.852491572	\\
0.02	1.854185525	\\
0.03	1.855863165	\\
0.04	1.857524315	\\
0.05	1.859168794	\\
0.06	1.860796425	\\
0.07	1.862323581	\\
0.08	1.86382252	\\
0.09	1.865303528	\\
0.1	1.866766423	\\
0.11	1.868211019	\\
0.12	1.869568052	\\
0.13	1.870877608	\\
0.14	1.872167754	\\
0.15	1.873438299	\\
0.16	1.874689054	\\
0.17	1.875820377	\\
0.18	1.876929639	\\
0.19	1.878017971	\\
0.2	1.879085175	\\
0.21	1.880062808	\\
0.22	1.880982936	\\
0.23	1.881880774	\\
0.24	1.882756122	\\
0.25	1.883537312	\\
0.26	1.884259944	\\
0.27	1.884958902	\\
0.28	1.885633981	\\
0.29	1.886177583	\\
0.3	1.88669426	\\
0.31	1.887185853	\\
0.32	1.887590024	\\
0.33	1.887918497	\\
0.34	1.88822067	\\
0.35	1.888454916	\\
0.36	1.888589241	\\
0.37	1.888696036	\\
0.38	1.888731132	\\
0.39	1.888665347	\\
0.4	1.888570795	\\
0.41	1.888378693	\\
0.42	1.888106831	\\
0.43	1.887804956	\\
0.44	1.887358713	\\
0.45	1.886874812	\\
0.46	1.886302955	\\
0.47	1.885633362	\\
0.48	1.884917007	\\
0.49	1.884058084	\\
0.5	1.883165842	\\
0.51	1.882127345	\\
0.52	1.881041146	\\
0.53	1.879820357	\\
0.54	1.878536641	\\
0.55	1.87711714	\\
0.56	1.875632395	\\
0.57	1.873998493	\\
0.58	1.872304742	\\
0.59	1.870445971	\\
0.6	1.868510884	\\
0.61	1.866441868	\\
0.62	1.864251736	\\
0.63	1.861969197	\\
0.64	1.859510959	\\
0.65	1.856950038	\\
0.66	1.854272923	\\
0.67	1.851421717	\\
0.68	1.848450279	\\
0.69	1.845351181	\\
0.7	1.842088173	\\
0.71	1.83867047	\\
0.72	1.835110212	\\
0.73	1.831401167	\\
0.74	1.827536963	\\
0.75	1.823485337	\\
0.76	1.819272511	\\
0.77	1.814893361	\\
0.78	1.8103431	\\
0.79	1.805617273	\\
0.8	1.800711752	\\
0.81	1.795622732	\\
0.82	1.79034673	\\
0.83	1.784880585	\\
0.84	1.779221455	\\
0.85	1.773366821	\\
0.86	1.767314488	\\
0.87	1.761062585	\\
0.88	1.754609572	\\
0.89	1.747954244	\\
0.9	1.741095733	\\
0.91	1.734033521	\\
0.92	1.726767439	\\
0.93	1.719297684	\\
0.94	1.711624819	\\
0.95	1.703749793	\\
0.96	1.695673944	\\
0.97	1.687399017	\\
0.98	1.678927176	\\
0.99	1.670257407	\\
1	1.661395017	\\
	};
	\addlegendentry{$E_{\mbox{\tiny trc}}(0)$}	
	
	\end{axis}
	
	\end{tikzpicture}
	\caption{Zero--rate error exponents as functions of $\alpha$ for $P=10$, $Q=\sigma_Z^2=1$. 
	The random coding exponent, $E_{\mbox{\tiny r}}(0)$ and the typical--code error exponent, $E_{\mbox{\tiny trc}}(0)$, are in solid and dashed lines, respectively.		
		The exponents $E_{\mbox{\tiny trc}}(0)$ and $E_{\mbox{\tiny ex}}(0)$ coincide.} \label{graph3}
\end{figure}
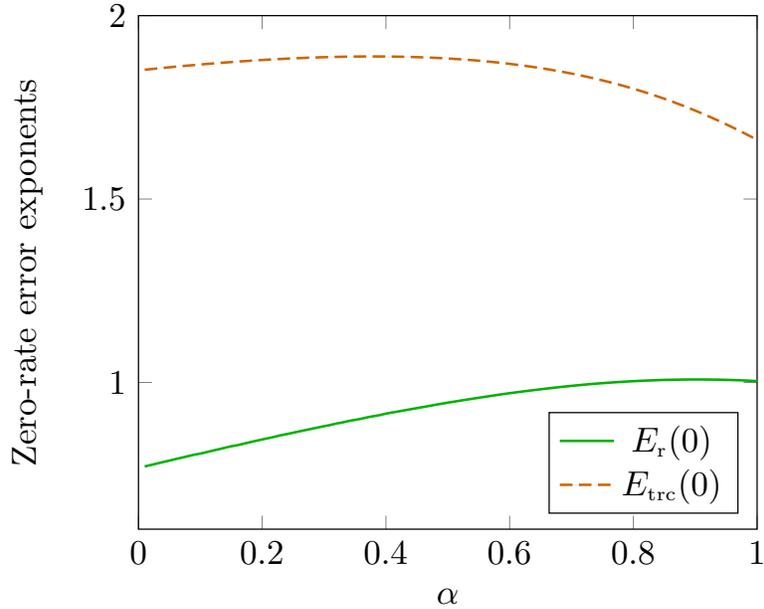

The random coding exponent, $E_{\mbox{\tiny r}}(R)$, 
the typical--code error exponent, $E_{\mbox{\tiny trc}}(R)$, and the expurgated error exponent,
$E_{\mbox{\tiny ex}}(R)$, are presented in Figure \ref{graph1}. As can be seen in Figure \ref{graph1}, $E_{\mbox{\tiny trc}}(R)$ and $E_{\mbox{\tiny ex}}(R)$ coincide at $R=0$, but $E_{\mbox{\tiny trc}}(R) < E_{\mbox{\tiny ex}}(R)$ at relatively positive low coding rates. At relatively high rates, the three exponent functions coincide.  
Figure \ref{graph2} presents zero-rate error exponents as a function of $Q$
for the DPC with the parameters $P=10$ and $\sigma_Z^2=1$, and with optimized $\alpha$ in the interval $[0,1]$. 
The two exponents $E_{\mbox{\tiny trc}}(0)$ and $E_{\mbox{\tiny ex}}(0)$ coincide.
Note that the random coding exponent
is almost a constant, independently of $Q$, in agreement with \cite{LW2006}. It turns out that for larger values of $Q$,
$E_{\mbox{\tiny r}}(0)$ is slightly increasing with $Q$, 
which may seem counter--intuitive. However, note that for $Q> 0$, even if we 
select $\alpha=1$ (which is sub-optimal), we 
completely cancel the interference (yielding $\bY=(\bU-\bS)+\bS+\bZ=\bU+\bZ$, 
yet the effective transmission power (the power of $\bU$) is $P+Q$ (rather than $P$), although the transmission power (of $\bX=\bU-\bS$) is still $P$.
On the other hand, there is the 
effective rate reduction of $I(U;S)=\frac{1}{2}\ln(1+\frac{Q}{P})$ (which does not exist when
$Q=0$). Still, the overall effect of increasing $Q$ might be positive.
Figure \ref{graph3} presents zero--rate error exponents as a function of $\alpha$ for the parameters $P=10$ and $Q=\sigma_Z^2=1$. 
The optimal values of $\alpha$ are different from each other: $\alpha_{\mbox{\tiny r}}^*\approx 0.90$,
$\alpha_{\mbox{\tiny trc}}^*=\alpha_{\mbox{\tiny ex}}^*\approx 0.38$.
The latter two are also different from the optimal, capacity--achieving $\alpha$, which is $\alpha_{\mbox{\tiny capacity}}^*=
P/(P+\sigma_Z^2)=10/11=0.9091$.

\section{Error Exponents of the Gel'fand-Pinsker Channel} \label{SEC_GP}


\subsection{Setting and Objectives} \label{SEC_GP_setup} 

Consider a state dependent DMC $W:~\calX\times\calS \to \calY$ with finite input, state and output alphabets $\calX$, $\calS$ and $\calY$, respectively. The $\calS$-valued state process $\{S_{t}\}_{t=1}^{\infty}$ is stationary and memoryless with known probability mass function $P_{S}$. The probability law of the DMC is specified by
\begin{align}
W(\by|\bx,\bs) = \prod_{t=1}^{n} W(y_{t}|x_{t},s_{t}),
\end{align}
where $\bx \in \calX^{n}$, $\bs \in \calS^{n}$, and $\by \in \calY^{n}$.

Let the coding rate $R$ be given. 
Let $\calU$ be a finite auxiliary alphabet. 
For a given type $Q_{S} \in \calP_{n}(\calS)$, let $Q_{U|S}(Q_{S})$ be any conditional type (to be optimized for every $Q_{S}$).
Furthermore, for any $Q_{S}$, let $Q_{X|SU}(Q_{S})$ be another conditional type (to be optimized as well). 
Let $\eps>0$ be given. For $Q=Q_{U|S} \times Q_{S}$, let $R(Q_{S}) = I_{Q}(S;U)+\eps$. 
For any $Q_{S} \in \calP_{n}(\calS)$, draw $e^{n(R+R(Q_{S}))}$ independent $n$-vectors, uniformly over the type class $\calT(Q_{U})$, where $Q_{U}$ is the $U$-marginal of $Q_{U|S} \times Q_{S}$. Partition these codewords into $M=e^{nR}$ bins, each one of size $M(Q_{S})=e^{nR(Q_{S})}$.    
Let us denote those bins as 
\begin{align}
\calB(Q_{S},m)=\{\bu_{Q_{S},m,1},\bu_{Q_{S},m,2},\ldots,\bu_{Q_{S},m,M(Q_{S})}\},~~~Q_{S}\in \calP_{n}(\calS),~~m \in \{1,2,\ldots,M\}.
\end{align} 
Note that this random-binning code construction is similar to the one in Subsection \ref{SEC_Objectives}, but here, we allow the binning rate to depend on the SI type, thus allowing for more degrees of freedom to improve the overall performance.

For a given message $m \in \{1,2,\ldots,M\}$ and a state vector $\bs \in \calS^{n}$, we choose the vector $\bu$ from the set $\calC(m,\bs) \dfn \calB(\hat{P}_{\bs},m) \cap \calT(Q_{U|S}|\bs)$ with equal probabilities. The probability that $\calC(m,\bs)$ is empty decays double-exponentially fast\footnote{This fact follows from Lemma \ref{lemmaK} in Subsection \ref{Sec_Preparatory_Steps}.}, hence we neglect this possible error event. 
For a given $\bs \in \calS^{n}$ and a selected vector $\bu$, we draw the codeword $\bx$ according to the uniform distribution over the conditional type class $\calT(Q_{X|SU}|\bs,\bu)$ and transmit it over the channel. 

Following the same rationale as in Subsection \ref{SEC_Objectives}, also here we implement a sub-optimal two-stage decoder, which first decodes for the $\bu$ codeword according to some metric that depends on the joint empirical distribution with the received vector $\by$, and afterwards, provides only the bin index that corresponds to the decoded $\bu$. 
More precisely, the decoder observes the received vector $\by$ and decodes for a bin index using the following generalized deterministic decoder:
\begin{align}
\label{DEF_GLD}
\hat{m}(\by)  =
\operatorname*{arg\,max}_{m \in \{1,2,\ldots,M\}} \left\{ \max_{Q_{S}\in \calP_{n}(\calS)} \max_{\bu \in \calB(Q_{S},m)} G(\hat{P}_{\bu\by},Q_{S}) \right\},
\end{align}
where $\hat{P}_{\bu\by}$ is the empirical distribution of $(\bu,\by)$, and $G(\cdot,\cdot)$ is a given continuous, real--valued functional of this empirical distribution and of another distribution $Q_{S} \in \calP_{n}(\calS)$.

For a given codebook, the probability of error is given by
\begin{align}
P_{\mbox{\tiny e}}(\calC_{n})
=\frac{1}{M}\sum_{m=1}^{M}
\sum_{\bs \in \calS^{n}} P(\bs)
\sum_{\by \in \calY^{n}} W(\by|\bx,\bs) \cdot \IND \{\hat{m}(\by) \neq m\}.
\end{align}
Our objectives here are identical to those defined in Subsection \ref{SEC_Objectives}.

\subsection{Main Results}

In order to present the random coding error exponent, define the exponent function
\begin{align} \label{RC_exp}
E_{\mbox{\tiny r}}(R) = \min_{Q_{S}} \max_{Q_{UX|S}} \min_{Q_{Y|SUX}} \{D(Q_{S}\|P_{S}) + &D(Q_{Y|SUX} \| W_{Y|XS}|Q_{UX|S} \times Q_{S}) \nn \\
&~+ [I_{Q}(U;Y) - I_{Q}(U;S) - R]_{+} \}.
\end{align}
Then, we have the following result, which is proved in Appendix C.
\begin{theorem} \label{THEOREM1}
	Consider the setting described in Subsection \ref{SEC_GP_setup}. \begin{enumerate}
		\item It holds that
		\begin{align}
		\lim_{n \to \infty} -\frac{1}{n} \ln \mathbb{E} [P_{\mbox{\tiny e}}(\calC_{n})] = E_{\mbox{\tiny r}}(R).
		\end{align}
		\item The universal decoder 
		\begin{align} \label{Uni_Dec}
		\hat{m}  =
		\operatorname*{arg\,max}_{m \in \{1,2,\ldots,M\}} \left\{ \max_{Q_{S}\in \calP_{n}(\calS)} \max_{\bu \in \calB(Q_{S},m)} [\hat{I}_{\bu\by}(U;Y) - R(Q_{S})] \right\}
		\end{align}
		achieves $E_{\mbox{\tiny r}}(R)$.
	\end{enumerate}  
\end{theorem}
\textbf{Discussion} An expression similar to \eqref{RC_exp} can be found in two previous works. First in \cite{SM2004}, where the coding technique is slightly different than ours and composed by two steps. First, the empirical distribution of the state sequence is transmitted to the receiver, and only then, the message is encoded similarly as in Subsection \ref{SEC_GP_setup}, by using a codebook, which is optimally designed for the empirical statistics of $\bs$. Since the decoder knows in which codebook to look at, it uses an ordinary MMI decoder to decode for a bin index. 
The final expression in \cite{SM2004} is given by the minimum between \eqref{RC_exp} and another expression, which is related to the first decoding phase of transmitting the statistics of $\bs$. Hence, our exponential error bound is at least as tight as the one in \cite{SM2004}.          
The expression in \eqref{RC_exp} appears also in \cite{Moulin2007}, where a similar codebook generation has already been encountered. Although the universal decoder \eqref{Uni_Dec} has already been proposed in \cite{Moulin2007}, its optimality has not been proved before.       

We continue by presenting a numerical example of the exact error exponent given in Theorem \ref{THEOREM1}. 
Let $\calS=\calU=\calX=\calY = \{0,1\}$ and $\pr\{S=0\}=1-\pr\{S=1\}=p \in [0,1]$. 
It is important to note that choosing $|\calU|=2$ may not be optimal, but we make this choice in order to keep the computational complexity relatively low.
Regarding the state-dependent channel, consider the following. If $S=0$, then $Y=X$ with probability one, i.e., the channel is clean. Otherwise, if $S=1$, the channel is stuck at 0: $\pr\{Y=0|X=0\}=1-\pr\{Y=1|X=0\}=1$ and $\pr\{Y=0|X=1\}=1-\pr\{Y=1|X=1\}=1$.     
The capacity of this channel is given by 
\begin{align}
C_{\mbox{\tiny GP}} = \max_{Q_{U|S}, X(U,S)} \{I(U;Y)-I(U;S)\} = p ~~ \text{[bits]},
\end{align}
which follows as a minor modification of \cite[pp.\ 178-180, Example 7.3]{El-Gamal}.

\begin{figure}[ht!]
	\centering
	\begin{tikzpicture}[scale=1.2]
	\begin{axis}[
	disabledatascaling,
	scaled x ticks=false,
	xticklabel style={/pgf/number format/fixed,
		/pgf/number format/precision=3},
	scaled y ticks=false,
	yticklabel style={/pgf/number format/fixed,
		/pgf/number format/precision=3},
	xlabel={$R$},
	ylabel={Error Exponents},
	xmin=0, xmax=0.5,
	ymin=0, ymax=0.45,
	xtick={0,0.1,0.2,0.3,0.4,0.5},
	legend pos=north east,
	]
	
	\addplot[smooth,color=black!30!green,thick]
	table[row sep=crcr] 
	{
0	4.31E-01	\\
0.01	4.21E-01	\\
0.02	4.11E-01	\\
0.03	4.01E-01	\\
0.04	3.91E-01	\\
0.05	3.81E-01	\\
0.06	3.71E-01	\\
0.07	3.61E-01	\\
0.08	3.51E-01	\\
0.09	3.41E-01	\\
0.1	3.31E-01	\\
0.11	3.21E-01	\\
0.12	3.11E-01	\\
0.13	3.01E-01	\\
0.14	2.91E-01	\\
0.15	2.81E-01	\\
0.16	2.71E-01	\\
0.17	2.61E-01	\\
0.18	2.51E-01	\\
0.19	0.240782866	\\
0.2	0.230782866	\\
0.21	0.220782866	\\
0.22	0.210782866	\\
0.23	2.01E-01	\\
0.24	0.190782866	\\
0.25	0.180782866	\\
0.26	0.170782866	\\
0.27	0.160782866	\\
0.28	0.150782866	\\
0.29	0.140782866	\\
0.3	0.130782866	\\
0.31	1.21E-01	\\
0.32	1.11E-01	\\
0.33	1.01E-01	\\
0.34	9.08E-02	\\
0.35	8.08E-02	\\
0.36	7.08E-02	\\
0.37	6.08E-02	\\
0.38	5.10E-02	\\
0.39	4.20E-02	\\
0.4	3.39E-02	\\
0.41	2.66E-02	\\
0.42	2.02E-02	\\
0.43	1.46E-02	\\
0.44	9.87E-03	\\
0.45	6.06E-03	\\
0.46	3.15E-03	\\
0.47	1.18E-03	\\
0.48	0.00015162	\\
0.49	5.55E-17	\\
0.5	5.55E-17	\\
	};
	\legend{}
	\addlegendentry{$E_{\mbox{\tiny r}}(R),~p=0.7$}

	\addplot[smooth,color=black!20!orange,thick,dash pattern={on 2pt off 1pt}]
table[row sep=crcr] 
{
0	2.88E-01	\\
0.01	2.78E-01	\\
0.02	2.68E-01	\\
0.03	2.58E-01	\\
0.04	2.48E-01	\\
0.05	2.38E-01	\\
0.06	2.28E-01	\\
0.07	2.18E-01	\\
0.08	2.08E-01	\\
0.09	1.98E-01	\\
0.1	1.88E-01	\\
0.11	1.78E-01	\\
0.12	1.68E-01	\\
0.13	1.58E-01	\\
0.14	1.48E-01	\\
0.15	1.38E-01	\\
0.16	1.28E-01	\\
0.17	1.18E-01	\\
0.18	1.08E-01	\\
0.19	9.77E-02	\\
0.2	8.77E-02	\\
0.21	7.77E-02	\\
0.22	6.77E-02	\\
0.23	5.77E-02	\\
0.24	4.81E-02	\\
0.25	3.94E-02	\\
0.26	3.16E-02	\\
0.27	2.46E-02	\\
0.28	1.86E-02	\\
0.29	0.013389611	\\
0.3	0.009059067	\\
0.31	5.59E-03	\\
0.32	2.95E-03	\\
0.33	1.15E-03	\\
0.34	1.84E-04	\\
0.35	-1.11E-16	\\
0.36	-1.11E-16	\\
0.37	-1.11E-16	\\
0.38	-1.11E-16	\\
0.39	-1.11E-16	\\
0.4	-1.11E-16	\\
0.41	-1.11E-16	\\
0.42	-1.11E-16	\\
0.43	-1.11E-16	\\
0.44	-1.11E-16	\\
0.45	-1.11E-16	\\
0.46	-1.11E-16	\\
0.47	-1.11E-16	\\
0.48	-1.11E-16	\\
0.49	-1.11E-16	\\
0.5	-1.11E-16	\\
};
\addlegendentry{$E_{\mbox{\tiny r}}(R),~p=0.5$}

	\addplot[smooth,color=black!20!purple,thick,dash pattern={on 3pt off 2pt}]
table[row sep=crcr] 
{
0	1.63E-01	\\
0.01	1.53E-01	\\
0.02	1.43E-01	\\
0.03	1.33E-01	\\
0.04	1.23E-01	\\
0.05	1.13E-01	\\
0.06	1.03E-01	\\
0.07	9.25E-02	\\
0.08	8.25E-02	\\
0.09	7.25E-02	\\
0.1	6.25E-02	\\
0.11	5.25E-02	\\
0.12	4.25E-02	\\
0.13	3.30E-02	\\
0.14	2.47E-02	\\
0.15	1.78E-02	\\
0.16	1.21E-02	\\
0.17	7.54E-03	\\
0.18	4.08E-03	\\
0.19	1.69E-03	\\
0.2	3.46E-04	\\
0.21	1.94E-16	\\
0.22	1.94E-16	\\
0.23	1.94E-16	\\
0.24	1.94E-16	\\
0.25	1.94E-16	\\
0.26	1.94E-16	\\
0.27	1.94E-16	\\
0.28	1.94E-16	\\
0.29	1.94E-16	\\
0.3	1.94E-16	\\
0.31	1.94E-16	\\
0.32	1.94E-16	\\
0.33	1.94E-16	\\
0.34	1.94E-16	\\
0.35	1.94E-16	\\
0.36	1.94E-16	\\
0.37	1.94E-16	\\
0.38	1.94E-16	\\
0.39	1.94E-16	\\
0.4	1.94E-16	\\
0.41	1.94E-16	\\
0.42	1.94E-16	\\
0.43	1.94E-16	\\
0.44	1.94E-16	\\
0.45	1.94E-16	\\
0.46	1.94E-16	\\
0.47	1.94E-16	\\
0.48	1.94E-16	\\
0.49	1.94E-16	\\
0.5	1.94E-16	\\
};
\addlegendentry{$E_{\mbox{\tiny r}}(R),~p=0.3$}	
	
	\end{axis}
	
	\end{tikzpicture}
	\caption{Random coding error exponents in the Gel'fand-Pinsker channel which is a clean channel when $S=0$ and a `stuck at 0' channel when $S=1$, for three different values of $p=\pr\{S=0\}$. 
	Not very surprising, the reliability of this channel is a monotone function of $p$.} \label{Fig:Error_Exponent}
\end{figure}
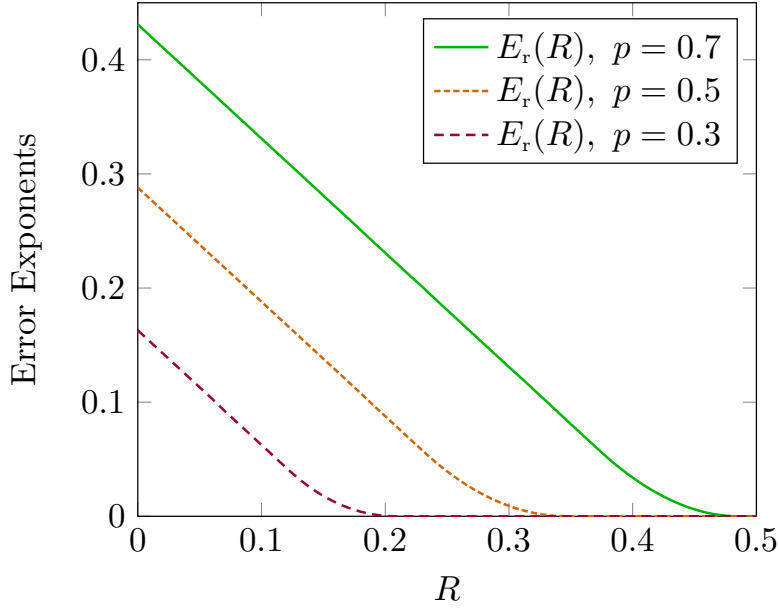

We plot the exponent function itself in Figure \ref{Fig:Error_Exponent} and find that its functional behavior is similar to that of the random coding error exponent in a DMC without random states; it decreases in an affine fashion for relatively low coding rates, and in a strictly convex fashion at the high coding rates. 
The reliability of this state-dependent channel obviously depends on the probability of $S=0$; the higher this probability is, the percentage of time the channel is clean is bigger, and the channel is more reliable, although the decoder is ignorant of the state realization. As can be seen in Figure \ref{Fig:Error_Exponent}, where we compare between $p=0.7$, $p=0.5$, and $p=0.3$, the reliability of the channel is highest at $p=0.7$ and lowest at $p=0.3$.

Following the studies in \cite{BargForney, MERHAV_TYPICAL, PRAD2014} on TRCs in ordinary channel coding, we claim that also in the Gel'fand-Pinsker Channel, the random coding error exponent, whose exact value is given in Theorem \ref{THEOREM1}, does not yield the true exponential behavior of the error probability of a randomly chosen code, since it is dominated by the relatively bad codes in the ensemble, rather than the channel noise, at least at low coding rates.
Due to the definition of the TRC exponent, the derivation of a single-letter expression is not as easy as in ordinary random coding (for example, see the proof in \cite[Section 5]{MERHAV_TYPICAL}), since the expectations over the randomness of the ensemble and over the randomness of the channel cannot be switched, which is one of the first steps in random coding analysis. 
Let us choose the universal decoding metric $g(Q_{UY},Q_{S}) = I_{Q}(U;Y) - R(Q_{S})$.
In order to present a lower bound on the error exponent of the TRC, define the expressions
\begin{align} \label{exp_Theta}
E_{0}(Q_{UU'SX},Q_{S'})
=\min_{\{Q_{Y|UU'SX}:~g(Q_{U'Y},Q_{S'}) \geq g(Q_{UY},Q_{S})\}} D(Q_{Y|UU'SX}\|W_{Y|SX}|Q_{UU'SX}),
\end{align}
\begin{align}
E_{1}(Q_{UU'S},Q_{S'})
= \min_{\{\bar{Q}_{X|UU'S}:~\bar{Q}_{X|US}=Q_{X|US}\}} [I_{\bar{Q}}(U';X|US)
+ E_{0}(Q_{UU'S} \times \bar{Q}_{X|UU'S},Q_{S'})],
\end{align}
and,
\begin{align}
E_{2}(Q_{UU'},Q_{S'})
= \min_{\{\tilde{Q}_{S|UU'}:~\tilde{Q}_{S}=Q_{S}\}} [I_{\tilde{Q}}(S;U'|U) + E_{1}(Q_{UU'} \times \tilde{Q}_{S|UU'},Q_{S'})].
\end{align} 
Denote the set $\calQ(Q_{U},Q_{U'})=\{\ddot{Q}_{UU'}:~\ddot{Q}_{U}=Q_{U},\ddot{Q}_{U'}=Q_{U'}\}$ and define the exponent function
\begin{align} \label{TRC_exp}
E_{\mbox{\tiny trc}}(R) 
&= \max_{\{Q_{UX|S}\}} \min_{Q_{S},Q_{S'}} \min_{\{\ddot{Q}_{UU'} \in \calQ(Q_{U},Q_{U'}):~I_{\ddot{Q}}(U;U') \leq 2R + R(Q_{S}) + R(Q_{S'})\}} \nn \\
&~~~~~~~~~~~[D(Q_{S}\|P_{S}) + E_{2}(\ddot{Q}_{UU'},Q_{S'}) + I_{\ddot{Q}}(U;U') - R - R(Q_{S'})].
\end{align}
Then, our second result is the following theorem, which is proved in Appendix D.
\begin{theorem} \label{THEOREM_TRC}
	Consider the setting described in Subsection \ref{SEC_GP_setup}. It holds that
	\begin{align}
	\liminf_{n \to \infty}  -\frac{1}{n} \mathbb{E} [\ln P_{\mbox{\tiny e}}(\calC_{n})] \geq E_{\mbox{\tiny trc}}(R).
	\end{align}  
\end{theorem}
Since our analysis in Appendix D amounts only to pairwise error events, the resulting exponent function is only a lower bound in general and is not tight at relatively high coding rates.   
Note that at high coding rates, the random coding error exponent of \eqref{RC_exp} provides the exact value of the error exponent of the TRC.

In ordinary channel coding, the random coding error exponent, and the error exponent of the TRC are both improved at relatively low coding rates by code expurgation. 
As we already seen in Figure \ref{graph1} above, this fact is also true for the DPC.
Upon using similar techniques as in \cite[Appendix A]{TM_UD}, which is an error exponent derivation of an expurgated code in ordinary channel coding, we are able to derive a bound which is tighter than $E_{\mbox{\tiny r}}(R)$ and $E_{\mbox{\tiny trc}}(R)$, at least at low coding rates. 
Let us define the exponent function
\begin{align} \label{EX_exp}
E_{\mbox{\tiny ex}}(R)
&= \max_{\{Q_{UX|S}\}} \min_{Q_{S},Q_{S'}} \min_{\{\ddot{Q}_{UU'} \in \calQ(Q_{U},Q_{U'}):~I_{\ddot{Q}}(U;U') \leq R + R(Q_{S}) + R(Q_{S'})\}} \nn \\
&~~~~~~~~~~~[D(Q_{S}\|P_{S}) + E_{2}(\ddot{Q}_{UU'},Q_{S'}) + I_{\ddot{Q}}(U;U') - R - R(Q_{S'})].
\end{align}
Then, our third result is the following theorem, which is proved in Appendix E.
\begin{theorem} \label{THEOREM_EX}
	Consider the setting described in Subsection \ref{SEC_GP_setup}.
	There exists a sequence of constant composition codes, $\{\calC_{n},~n=1,2,\dotsc\}$, such that
	\begin{align}
	\liminf_{n \to \infty} - \frac{1}{n} \ln \max_{m} P_{\mbox{\tiny e}|m}(\calC_{n}) \geq E_{\mbox{\tiny ex}}(R).
	\end{align}
\end{theorem}


\section{Proof of Theorem \ref{Theorem_DPC}}
\label{Sec_Proof_DPC}

\subsection{Preparatory Steps} \label{Sec_Preparatory_Steps}
Before moving on to the actual derivations of the various error exponent functions,
two basic preparatory steps will prove useful. The first is associated with certain helpful properties
that a randomly chosen code in the ensemble satisfies with an overwhelmingly high probability,
and the second is associated with the structure of the joint 
distribution of the various variables in the coded communication system under discussion.\\

\noindent
{\bf 1.~Two properties of good codes}\\
There are two desired properties that are associated with a randomly chosen code in our ensemble,
with an extremely high probability, and we henceforth refer to codes with those properties
as {\em good codes}. 
Roughly speaking, a good code has the following two properties
at the same time: (i) 
$K_n(s,m)$ is close to $e^{n\Delta}$ for all $\bs$ and $m$, and (ii) For every given $\by$, 
there exists at least one codeword, $\bu_k$, such that $\hat{\rho}^2(\bu_k,\by)$ is essentially 
at least as large as $1-\exp[-2(R+I_{US})]$.
These two features will be pivotal in our analysis.

To state properties (i) and (ii) more formally, we assert the two following corresponding lemmas, 
whose proofs appear in Appendix B.
\begin{lemma}
	\label{lemmaK}
	Let $\epsilon > 0$ ($\epsilon\ll\Delta$) be arbitrarily small
	and let $\calG_1^n$ be the subset of all codes such that
	\begin{equation}
	e^{n(\Delta-\epsilon)}\le K_n(\bs,m)\le e^{n(\Delta+\epsilon)} 
	\end{equation}
	for all $m\in\{0,1,\ldots,M-1\}$ and $\bs\in\calS(\sqrt{n\hat{Q}})\dfn\{\bs:~\|\bs\|^2=n\hat{Q}\}$. 
	Then, a randomly selected code falls in $\calG_1^n$ with probability that tends to unity double-exponentially
	as a function of $n$.
\end{lemma}

\begin{lemma}
	\label{lemmaRHO}
	Let $\epsilon > 0$ be arbitrarily small
	and let $\calG_2^n$ be the subset of all codes such that
	for all $k,k'\in\{0,1,\ldots,M-1\}$, $k'\ne k$,
	\begin{equation}
	\max_{\tilde{k}\ne k,k'}\hat{\rho}^2(\bu_{\tilde{k}},\by)\ge 1-\exp[-2(R+I_{US}-\epsilon)],
	\end{equation}
	for all $\by\in\reals^n$.
	Then, a randomly selected code falls in $\calG_2^n$ with probability that tends to unity double-exponentially
	as a function of $n$.
\end{lemma}

Finally, a good code is in $\calG^n=\calG_1^n\cap\calG_2^n$, which still holds
a probability that tends to unity double-exponentially rapidly as $n$ tends to infinity.\\

\noindent
{\bf 2.~On the structure of the joint probability disribution of $(k,\bs,\by)$}

Since our decoder, first decodes the index $k$ of the auxuliary codeword, $\bu_k$ ($0\le k\le M-1$) and
then decodes the message to be the index $m$ 
of the bin to which $k$ belongs (see eqs.\ (\ref{decodek}), (\ref{decodem})), 
an obvious upper bound to the probability
of error in estimating $m$ is the probability of error in estimating $k$, since the latter also counts errors
associated with incorrect values of $k$ that still belong to the correct bin, $m$. But the gap is negligible since
the number of incorrect codewords in the correct bin is only $M_2-1$, which is of a smaller exponential order than
$M-1$. We therefore henceforth approach the error probability analysis in the same way as if we had to assess the
probability of error in decoding $k$. To this end, it is instrumental 
to characterize the joint distribution of $k$ with
other entities that play a role in the coded communication system, namely, $\bs$ and $\by$.

We first observe that for a good code, the probability distribution of the index $k$ is essentially uniform.
To see why this is true, consider first the probability of $k$ given the message $m$, where
$m=\lfloor k/M_2\rfloor$. 
\begin{eqnarray}
P(k|m)&=&\int_{\reals^n}P(\bs)\cdot 
e^{-n\Delta} \IND \{\bu_k\in\calC_m\cap\calT(\bU|\bs)\}\cdot\mbox{d}\bs\nonumber\\
&=&\int_{\calT(\bS|\bu_k)}P(\bs)e^{-n\Delta}\mbox{d}\bs\nonumber\\
&=&e^{-n\Delta}\int_{\calT(\bS|\bu_k)}\frac{\mbox{d}\bs}{\mbox{Surf}\{\calS(\sqrt{n\hQ})\}},
\end{eqnarray}
where $\calT(\bS|\bu_k)=\{\bs:~\bu_k\in\calT(\bU|\bs)\}$. 
Similarly as in the proof of Lemma \ref{lemmaRHO}, the last integral is easily seen to be of the
exponential order of $\exp\{\frac{n}{2}\ln(1-\rho_0^2)\}=e^{-nI_{US}}$, and so,
\begin{equation}
P(k|m)\exe e^{-n(I_{US}+\Delta)}=\frac{1}{M_2},~~~\mbox{whenever}~m=\left\lfloor \frac{k}{M_2}\right\rfloor
\end{equation}
and therefore,
\begin{equation}
P(k)\exe\pr\left\{m=\left\lfloor 
\frac{k}{M_2}\right\rfloor\right\}\cdot P(k|m)=\frac{1}{M_1}\cdot\frac{1}{M_2}=\frac{1}{M},~~~~k=0,1,\ldots,M-1
\end{equation}
where we have assumed that the prior distribution over the various messages, $\{m\}$, is uniform.

For a good code, the joint distribution of $(k,\bs,\by)$ is given by
\begin{eqnarray}
P(k,\bs,\by)&=&P(k)P(\bs|\bu_k)P(\by|\bu_k,\bs)\nonumber\\
&\exe&\frac{1}{M}\cdot P(\bs|\bu_k)
P(\by|\bu_k,\bs),
\end{eqnarray}
where $P(\by|\bu_k,\bs)=\calN\left(\bu_k+(1-\alpha_{\bs})\bs,\sigma_Z^2I_n\right)$ is the additive white
Gaussian noise channel, with input $\bx+\bs=\bu_k+(1-\alpha_{\bs})\bs$ and noise variance, $\sigma_Z^2$,
and
\begin{eqnarray}
P(\bs|\bu_k)&=&\frac{P(\bs)\cdot P(\bu_k|\bs)}{\int_{\reals^n}P(\bs')\cdot 
	P(\bu_k|\bs')\mbox{d}\bs'}\nonumber\\
&=&\frac{P(\bs)\cdot e^{-n\Delta}\IND\{\bu_k\in\calC_m\cap\calT(\bU|\bs)\}}
{\int_{\reals^n}P(\bs') e^{-n\Delta}\IND\{\bu_k\in\calC_m\cap\calT(\bU|\bs')\}\mbox{d}\bs'}\nonumber\\
&=&\frac{[1/\mbox{Surf}\{\calS(\sqrt{n\hQ})\}]\IND\{\bu_k\in\calC_m\cap\calT(\bU|\bs)\}}
{\int_{\calT(\bS|\bu_k)}[1/\mbox{Surf}\{\calS(\sqrt{n\hQ})\}]\mbox{d}\bs'}\nonumber\\
&=&\frac{1}{\int_{\calT(\bS|\bu_k)}\mbox{d}\bs'}\nonumber\\
&\exe&\frac{1}{\mbox{Vol}\{\calT(\bS|\bu_k)\}},
\end{eqnarray}
where $\mbox{Vol}\{\calT(\bS|\bu_k)\}$ is the volume of $\calT(\bS|\bu_k)$.

As a consequence of these relations, we can now present the error probability (in $k$) as follows.
\begin{eqnarray}
P_{\mbox{\tiny e}}(\calC)&=&\frac{1}{M}\sum_{k=0}^{M-1}
\int_{\calT(\bS|\bu_k)}\frac{\mbox{d}\bs}{\mbox{Vol}\{\calT(\bs|\bu_k)\}}
\int_{\Lambda_k^{\mbox{\tiny c}}}P(\by|\bu_k,\bs)\mbox{d}\by\nonumber\\
&=&\frac{1}{M}\sum_{k=0}^{M-1}\int_{\Lambda_k^{\mbox{\tiny c}}}\left[
\int_{\calT(\bS|\bu_k)}\frac{P(\by|\bu_k,\bs)\mbox{d}\bs}{\mbox{Vol}\{\calT(\bs|\bu_k)\}}\right]
\mbox{d}\by\nonumber\\
&\dfn&\frac{1}{M}\sum_{k=0}^{M-1}\int_{\Lambda_k^{\mbox{\tiny c}}}P(\by|\bu_k)\mbox{d}\by,
\end{eqnarray}
where $\Lambda_k^{\mbox{\tiny c}}$ is the complement to the decision region in favor of $\bu_k$.
This is an ordinary expression of the probability of error, associated with a codebook $\calC=\{\bu_k\}$
over the effective channel
\begin{equation}
P(\by|\bu_k)=\int_{\calT(\bS|\bu_k)}\frac{P(\by|\bu_k,\bs)\mbox{d}\bs}{\mbox{Vol}\{\calT(\bs|\bu_k)\}}.
\end{equation}
According to this effective channel, $\by$ depends on $\bu_k$ only 
via the empirical correlation coefficient, $\hat{\rho}(\bu_k,\by)$. Therefore, the
optimal decoding metric also depends on $(\bu_k,\by)$ only via $\hat{\rho}(\bu_k,\by)$ (in other words,
$\hat{\rho}(\bu_k,\by)$ serves as sufficient statistics).
It follows from the results of \cite{MERHAV2013} (see in particular, Example 2 therein) that at least for the random coding error exponent, the MMI decoder is asymptotically optimal in the sense of achieving the same error exponent as the maximum likelihood decoder.

\subsection{Derivation}

Considering the MMI decoder, which is equivalent to the maximum absolute correlation decoder,
that chooses the index $k$ with maximum $|\bu_k^T\by|$, we have:
\begin{align}
P_{\mbox{\tiny e}}(\calC_n)
&\le \frac{1}{M}\sum_{k=0}^{M-1}\sum_{k'\ne k}\int_{\reals^n}P(\bs)\mbox{d}\bs
\int_{\reals^n}P(\by|\bu_k,\bs)\times \nn \\  &~~~~~~\IND\left\{|\bu_{k'}^T\by|\ge\max\left\{|\bu_k^T\by|,
\max_{\tilde{k}\ne k,k'}|\bu_{\tilde{k}}^T\by|\right\}\right\}\mbox{d}\by.
\end{align}
Our good codes (in $\calG_n$) satisfy
\begin{equation}
\forall k,k',\by,~~
\max_{\tilde{k}\ne k,k'}|\bu_{\tilde{k}}^T\by|\ge \sqrt{n(P+\alpha^2Q)\|\by\|^2}\sqrt{1-e^{-2(R_1-\epsilon)}},
\end{equation}
where $R_1=R+I_{US}$.
Therefore,
we can safely further bound the error probability of every $\calC_n\in\calG_n$ according to
\begin{align}
P_{\mbox{\tiny e}}(\calC_n)
\label{refa0}
&\lexe
\frac{1}{M}\sum_{k=0}^{M-1}\sum_{k'\ne
	k}\int_{\reals^n}P(\bs)\mbox{d}\bs\int_{\reals^n}P(\by|\bu_k,\bs)\times\nonumber\\
&~~~~~~\IND\left\{|\bu_{k'}^T\by|\ge\max\{|\bu_k^T\by|,\sqrt{n(P+\alpha^2Q)\|\by\|^2}\sqrt{1-e^{-2(R_1-\epsilon)}}\}
\right\}\mbox{d}\by \\
\label{refa1}
&= \frac{1}{M}\sum_{k=0}^{M-1}\sum_{k'\ne k}\int_{\reals^n}P(\bs)\mbox{d}\bs\int_{\reals^n} P(\by|\bu_k,\bs)\times\nonumber\\
&~~~~~~\IND\left\{
|\bu_{k'}^T\by|\ge|\bu_k^T\by|,|\bu_{k'}^T\by|\ge\sqrt{n(P+\alpha^2Q)\|\by\|^2}\sqrt{1-e^{-2(R_1-\epsilon)}}
\right\} \mbox{d}\by,
\end{align}
where \eqref{refa1} is due to the simple identity $\IND\{a\geq \max\{b,c\}\} = \IND\{a\geq b, a\geq c\}$. 
For the sake of notational simplicity, 
we henceforth use the following alternative notation: $\bu=\bu_k$, $\bv=\bu_{k'}$,
and $\bw=\bu+(1-\alpha_{\bs})\bs$. Now,
\begin{align}
I(\bs)
\label{refa2}
&\dfn(2\pi\sigma^2)^{-n/2}\int_{\reals^n}\mbox{d}\by\cdot
\exp\left\{-\frac{1}{2\sigma^2}\|\by-\bw\|^2\right\}
\cdot \IND\left\{
|\bv^T\by|\ge|\bu^T\by|,|\bv^T\by|\ge\sqrt{nW\|\by\|^2z}
\right\} \\
\label{refa3}
&\le\frac{\exp\{-\|\bw\|^2/2\sigma^2\}}{(2\pi\sigma^2)^{n/2}}
\int_{\reals^n}\mbox{d}\by\cdot\exp\left\{\frac{\by^T\bw}{\sigma^2}-
\frac{\|\by\|^2}{2\sigma^2}\right\}\times\nonumber\\
&~~~\exp\left\{\frac{\theta}{2}\by^T(\bv\bv^T-\bu\bu^T)\by+
\frac{\zeta}{2}\by^T(\bv\bv^T-nzW\cdot I_n)\by\right\} \\
\label{refa4}
&=\frac{\exp\{-\|\bw\|^2/2\sigma^2\}}{(2\pi\sigma^2)^{n/2}}
\int_{\reals^n}\mbox{d}\by\cdot\exp\left\{\frac{\by^T\bw}{\sigma^2}-\frac{1}{2}\by^T\Gamma\by\right\},
\end{align}
where \eqref{refa2} follows from the definitions of $W$ and $z$, $\theta>0$ and $\zeta>0$ in \eqref{refa3} are Chernoff parameters, and $\Gamma$ in \eqref{refa4} is defined by
\begin{eqnarray}
\Gamma&=&\left(\frac{1}{\sigma^2}+\zeta nzW\right)\cdot I_n+\theta\bu\bu^T-(\theta+\zeta)\bv\bv^T\nonumber\\
\label{refa7}
&=&\frac{1}{\sigma^2}\left[(1+\zeta n\sigma^2zW)\cdot I_n+\sigma^2\theta\bu\bu^T-
\sigma^2(\theta+\zeta)\bv\bv^T\right].
\end{eqnarray}
We now use the following identity
\begin{equation}
\frac{1}{(2\pi)^{n/2}|\Lambda|^{1/2}}\int_{\reals^n}\mbox{d}\by 
\exp\left\{\by^T\bomega-\frac{1}{2}\by^T\Lambda^{-1}\by\right\}=
\exp\left\{\frac{1}{2}\bomega^T\Lambda\bomega\right\},
\end{equation}
and so, continuing from \eqref{refa4},
\begin{align}
I(\bs)
&\le\frac{\exp\{-\|\bw\|^2/2\sigma^2\}}{(2\pi\sigma^2)^{n/2}}
\int_{\reals^n}\mbox{d}\by\cdot\exp\left\{\frac{\by^T\bw}{\sigma^2}-\frac{1}{2}\by^T\Gamma\by\right\}\\
&=\frac{\exp\{-\|\bw\|^2/2\sigma^2\}}{\sigma^n|\Gamma|^{1/2}}
\exp\left\{\frac{1}{2\sigma^4}\bw^T\Gamma^{-1}\bw\right\}\\
\label{refa5}
&=\frac{\exp\{-\|\bw\|^2/2\sigma^2\}}{\sigma^n|\Gamma|^{1/2}}
\exp\left\{\frac{1}{2\sigma^2}\bw^T[(1+\zeta n\sigma^2zW)I_n+
\sigma^2\theta\bu\bu^T-\sigma^2(\theta+\zeta)\bv\bv^T]^{-1}\bw\right\} \\
\label{refa6}
&=\frac{\exp\{-\|\bw\|^2/2\sigma^2\}}{|\mu I_n+t\bu\bu^T-r\bv\bv^T|^{1/2}}
\exp\left\{\frac{1}{2\sigma^2}\bw^T(\mu I_n+t\bu\bu^T-r\bv\bv^T)^{-1}\bw\right\},
\end{align}
where \eqref{refa5} follows by substituting \eqref{refa7}, and in \eqref{refa6} we have denoted $\mu=1+\zeta n\sigma^2zW$,
$t=\sigma^2\theta$, and $r=\sigma^2(\theta+\zeta)$.
We now have to find both the inverse and the determinant of $\mu I_n+t\bu\bu^T-r\bv\bv^T$,
where we use the fact that $\bu^T\bu=\bv^T\bv=nW$ and $\bu^T\bv=n\rho W$.
Beginning from the inverse, we invoke the matrix inversion lemma, asserting that for given matrices,
$A$, $B$, $C$, and $D$, of dimensions, $n\times n$, $n\times k$, $k\times k$, and $k\times n$, respectively
($k\le n$),
\begin{equation}
(A+BCD)^{-1}=A^{-1}-A^{-1}B(C^{-1}+DA^{-1}B)^{-1}DA^{-1}.
\end{equation}
Setting $k=2$, $A=\mu I_n$, $B=[t\bu~~r\bv]$, $C=I_2$, $D=[\bu~~-\bv]^T$, $p=nWt$ and $q=nWr$, we arrive at
\begin{align}
(\mu I_n+t\bu\bu^T-r\bv\bv^T)^{-1}
&=\frac{1}{\mu}I_n-\frac{t(\mu-q)\bu\bu^T-r(\mu+p)\bv\bv^T+\rho qt(\bu\bv^T+\bv\bu^T)}
{\mu[(\mu+p)(\mu-q)+\rho^2pq]} \\
\label{refa8}
&=\frac{1}{\mu}I_n-\frac{\mu t\bu\bu^T-\mu r\bv\bv^T-qt[\bu\bu^T+\bv\bv^T-\rho(\bu\bv^T+\bv\bu^T)]}
{\mu[(\mu+p)(\mu-q)+\rho^2pq]},
\end{align}
where we have used the fact that $pr=qt$.
To find the determinant of $\mu I_n+t\bu\bu^T-r\bv\bv^T$, we find the eigenvalues and take their product. 
First, observe that all $n-2$ linearly independent
vectors that are orthogonal to both $\bu$ and $\bv$ are eigenvectors pertaining to the 
eigenvalue $\mu$. The two remaining eigenvalues
correspond to two linearly independent vectors that lie 
in the subspace spanned by $\bu$ and $\bv$. Let us denote
$\br=a\bu+b\bv$. Then for $\br$ to be an eigenvector, 
\begin{eqnarray}
(\mu I_n+t\bu\bu^T-r\bv\bv^T)\br&=&
(\mu I_n+t\bu\bu^T-r\bv\bv^T)(a\bu+b\bv)\\
&=&\mu a\bu+\mu b\bv+ap\bu-a\rho q\bv+b\rho p\bu-bq\bv\\
&=&[\mu a+p(a+\rho b)]\bu+[\mu b-q(\rho a+b)]\bv.
\end{eqnarray}
The resulting eigenvalues are therefore the eigenvalues of the $2\times 2$ matrix
\begin{align}
\left(\begin{array}{cc}
\mu+p & \rho p\\
-\rho q & \mu-q\end{array}\right),
\end{align}
which are readily found to be $\lambda=\mu+(p-q)/2\pm \sqrt{(p+q)^2-4\rho^2pq}/2$.
Thus,
\begin{equation}
\label{refa9}
|\mu I_n+t\bu\bu^T-r\bv\bv^T|^{1/2}=\mu^{n/2-1}\sqrt{\mu^2+\mu(p-q)-pq(1-\rho^2)}.
\end{equation}
Now, substituting \eqref{refa8} and \eqref{refa9} back into \eqref{refa6} yields that 
\begin{align}
I(\bs)
&\le\frac{1}{\mu^{n/2-1}\sqrt{\mu^2+\mu(p-q)-pq(1-\rho^2)}}\times\nonumber\\
&~~\exp\bigg(-\frac{1}{2\sigma^2\mu}\bigg[n(\mu-1)[W+(1-\alpha^2)Q]+
\frac{\mu tA^2-\mu rB^2-qt(A^2-2\rho AB+B^2)}{(\mu+p)(\mu-q)+\rho^2pq}\bigg]\bigg) \\
\label{refa12}
&=\frac{1}{[1+(q-p)z]^{n/2-1}\sqrt{[1+(1-z)p+qz][1-
		pz-q(1-z)]+\rho^2pq}}\times\nonumber\\
&~~\exp\bigg(-\frac{1}{2\sigma^2\mu}\bigg[n(q-p)z[W+(1-\alpha^2)Q] \nn \\
&~~~~~~~+ \frac{[1+(q-p)z](pA^2-qB^2)-pq(A^2-2\rho AB+B^2)
}{nW([1+p(1-z)+qz][1-pz-q(1-z)]+\rho^2pq)}\bigg]\bigg),
\end{align}
where we have used the relations, $t=p/nW$, $r=q/nW$, $\mu=1+(q-p)z$ ($q\ge p$), and where
\begin{eqnarray}
A&=&\bw^T\bu\\
&=&(1-\hat{\alpha})\bs^T\bu+nW\\
&=&\hat{\alpha}(1-\hat{\alpha})\bs^T\bs+nW\\
&=&n[\alpha(\sqrt{\hat{Q}}-\alpha\sqrt{Q})\sqrt{Q}+W]\\
\label{refa10}
&\dfn&n\cdot a
\end{eqnarray}
and 
\begin{eqnarray}
B&=&\bw^T\bv \\
&=&(1-\hat{\alpha})\bs^T\bv+\rho nW \\
&=&(1-\hat{\alpha})\varrho\sqrt{nW}\|\bs\|+\rho nW\\
&=&n[(\sqrt{\hat{Q}}-\alpha\sqrt{Q})\sqrt{W}\varrho+\rho W] \\
\label{refa11}
&\dfn&n\cdot b,
\end{eqnarray}
with $\varrho$ designating the empirical correlation coefficient between $\bs$ and $\bv$.
Using both \eqref{refa10} and \eqref{refa11} in \eqref{refa12} and neglecting the factor in \eqref{refa12} that is independent of $n$, we have that
\begin{eqnarray}
I(\bs)&\lexe&\exp\bigg\{-n\bigg(\frac{1}{2}\ln[1+(q-p)z]+\frac{1}{2\sigma^2[1+(q-p)z]}\bigg[
(q-p)z[W+(1-\alpha^2)Q]+\nonumber\\
& &\frac{[1+(q-p)z](pa^2-qb^2)-pq(a^2-2\rho ab+b^2)
}{W([1+p(1-z)+qz][1-pz-q(1-z)]+\rho^2pq)}\bigg]\bigg)\bigg\} \\
&\dfn&e^{-nE(p,q,\hat{\sigma}_s^2,\rho,\varrho)},
\end{eqnarray}
where the dependence of $E(p,q,\hat{\sigma}_s^2,\rho,\varrho)$ on $\varrho$ is through $b$, and
where $E(p,q,\hat{\sigma}_s^2,\rho,\varrho)$ should be maximized over the set
\begin{align}
\calA=\{(p,q):~q\ge p\ge 0,~[1+p(1-z)+qz][1-pz-q(1-z)]+\rho^2pq> 0\},
\end{align}
resulting in
\begin{align}
E(\hat{\sigma}_s^2,\rho,\varrho)=\sup_{(p,q)\in\calA}E(p,q,\hat{\sigma}_s^2,\rho,\varrho).
\end{align}

As for the simpler bound in \eqref{Simp1}-\eqref{Simp2},
we now choose $q=p$ (amounting to simple union-bound analysis), which simplifies to
\begin{eqnarray}
\label{refa13}
E(q,q,\hat{\sigma}_s^2,\rho,\varrho)&=&
\frac{q(a^2-b^2)-q^2(a^2-2\rho ab+b^2)
}{2\sigma^2W[1-q^2(1-\rho^2)]},
\end{eqnarray}
and the supremum over $q$ should be taken in the range $q\in[0,1/\sqrt{1-\rho^2})$.
Alternatively, defining $\tau=q\sqrt{1-\rho^2}$, then \eqref{refa13} is equivalent to 
\begin{align}
E(\tau,\hat{\sigma}_s^2,\rho,\varrho)=\frac{1}{2\sigma^2W\sqrt{1-\rho^2}}\cdot\frac{g\tau-h\tau^2}{1-\tau^2},
\end{align}
which should be maximized over $\tau\in[0,1)$,
where $g=a^2-b^2$ and
\begin{align}
h=\frac{a^2-2\rho ab+b^2}{\sqrt{1-\rho^2}}.
\end{align}
Then,
\begin{equation}
\sup_{0\le\tau < 1}\frac{g\tau-h\tau^2}{1-\tau^2}=T(g,h)\dfn\left\{\begin{array}{ll}
0 & g\le 0\\
\frac{g^2}{2(h+\sqrt{h^2-g^2})} & h\ge g > 0\\
\infty & g > h > 0
\end{array}\right.
\end{equation}
It follows that
\begin{equation}
I(\bs)\lexe\exp\left\{-\frac{n}{2W\sigma^2\sqrt{1-\rho^2}}\cdot 
T\left(a^2-b^2,\frac{a^2-2\rho ab+b^2}{\sqrt{1-\rho^2}}\right)\right\}.
\end{equation}
Note that for a given $\hat{Q}=\|\bs\|^2/n$, 
and a given correlation coefficient, $\rho$, between $\bu$ and $\bv$, 
the variable $a$ is a deterministic constant and
$b$ depends only on the empirical correlation coefficient, $\varrho$. Let us then denote the simplified exponent for given
$\hat{\sigma}_s^2$, $\rho$, and $\varrho$, by 
\begin{equation}
E_{\mbox{\tiny sim}}(\hat{\sigma}_s^2,\rho,\varrho) =\frac{1}{2W\sigma^2\sqrt{1-\rho^2}}\cdot
T\left(a^2-b^2,\frac{a^2-2\rho ab+b^2}{\sqrt{1-\rho^2}}\right).
\end{equation}
The probability that $\bs$ would have
an empirical correlation coefficient, $\varrho$, with $\bv$ 
given that it has empirical correlation $\rho_0$ with $\bu$
is exponentially 
\begin{eqnarray}
P(\calT(\bs|\bu,\bv)|\bs\in\calT(\bs|\bu))&=&\frac{\mbox{Vol}\{\calT(\bs|\bu,\bv)\}}
{\mbox{Vol}\{\calT(\bs|\bu)\}}\\
\label{ToRef0}
&\exe&\frac{(2\pi e \sigma_s^2[1-(\rho_0^2+\varrho^2-2\rho\rho_0\varrho)/(1-\rho^2)])^{n/2}}
{(2\pi e \sigma_s^2[1-\rho_0^2])^{n/2}} \\
&\dfn& e^{-nZ(\rho,\varrho)},
\end{eqnarray}
where \eqref{ToRef0} is obtained similarly as in \cite{ArikanMerhav1998} and
\begin{eqnarray}
Z(\rho,\varrho)
&=&\frac{1}{2}\ln\left[\frac{1-\rho_0^2}
{1-(\rho_0^2+\varrho^2-2\rho\rho_0\varrho)/(1-\rho^2)}\right] \\
\label{refa14}
&=&\frac{1}{2}\ln\left[\frac{1}
{1-(\rho_0^2+\varrho^2-2\rho\rho_0\varrho)/(1-\rho^2)}\right]-
\frac{1}{2}\ln\left(1+\frac{\alpha^2Q}{P}\right) \\
&=&\frac{1}{2}\ln\left[\frac{1}
{1-\rho^2-\rho_0^2-\varrho^2+2\rho\rho_0\varrho}\right]+\frac{1}{2}\ln(1-\rho^2)-
\frac{1}{2}\ln\left(1+\frac{\alpha^2Q}{P}\right) \\
&\dfn&L(\rho,\varrho)+\frac{1}{2}\ln(1-\rho^2)-
\frac{1}{2}\ln\left(1+\frac{\alpha^2Q}{P}\right),
\end{eqnarray}
where \eqref{refa14} follows from \eqref{DEF_rho0}.
Thus, the overall exponent of a pairwise error is
\begin{eqnarray}
E(\hat{\sigma}_s^2,\rho)&=&\inf_{\{\varrho\in \calP(\rho,\rho_0)\}}\left[
E(\hat{\sigma}_s^2,\rho,\varrho)+Z(\rho,\varrho)\right]\\
&=&\inf_{\{\varrho\in\calP(\rho,\rho_0)\}}\left[
E(\hat{\sigma}_s^2,\rho,\varrho)+L(\rho,\varrho)\right]+\frac{1}{2}\ln(1-\rho^2)-
\frac{1}{2}\ln\left(1+\frac{\alpha^2Q}{P}\right)\\
&\dfn&\hat{E}(\hat{\sigma}_s^2,\rho)+\frac{1}{2}\ln(1-\rho^2)-\frac{1}{2}\ln\left(1+\frac{\alpha^2Q}{P}\right),
\end{eqnarray}
where $\calP(\rho,\rho_0)$ is the set of values of $\varrho$ such that the matrix
\begin{align}
\left(\begin{array}{ccc}
1 & \rho_0 & \rho\\
\rho_0 & 1 & \varrho\\
\rho & \varrho & 1\end{array}\right)
\end{align}
is non--negative definite.
Finally, we have the following exponents
\begin{eqnarray}
E_{\mbox{\tiny r}}(R,\hat{\sigma}_s^2)&=&\inf_{\{\rho:~|\rho| < 1\}}\left[E(\hat{\sigma}_s^2,\rho)
+\frac{1}{2}\ln\frac{1}{1-\rho^2}\right]-
\frac{1}{2}\ln\left(1+\frac{\alpha^2Q}{P}\right)-R\\
&=&\inf_{\{\rho:~|\rho| < 1\}}\hat{E}(\hat{\sigma}_s^2,\rho)-\ln\left(1+\frac{\alpha^2Q}{P}\right)-R.
\end{eqnarray}
Similarly,
\begin{eqnarray}
\label{TRCexponent}
E_{\mbox{\tiny trc}}(R,\hat{\sigma}_s^2)&=
&\inf_{\{\rho:~\rho^2 < 1-(P/W)^2e^{-4R}\}}\hat{E}(\hat{\sigma}_s^2,\rho)
-\ln\left(1+\frac{\alpha^2Q}{P}\right)-R\\
\label{EXexponent}
E_{\mbox{\tiny ex}}(R,\hat{\sigma}_s^2)&=&\inf_{\{\rho:~\rho^2 < 1-(P/W)^2e^{-2R}\}}\hat{E}(\hat{\sigma}_s^2,\rho)
-\ln\left(1+\frac{\alpha^2Q}{P}\right)-R.
\end{eqnarray}
In order to understand the constraint in \eqref{TRCexponent}, consider the following. The error probability is upper-bounded by
\begin{align}
P_{\mbox{\tiny e}}(\calC_n)
&\lexe
\frac{1}{M}\sum_{k=0}^{M-1}\sum_{k'\ne k} \exp\left\{-n E(\hat{\sigma}_s^2,\rho_{\bu_{k}\bu_{k'}}) \right\} \\
&\doteq \sum_{\rho_{UU'}} M(\rho_{UU'}) \exp\left\{-n [E(\hat{\sigma}_s^2,\rho_{UU'})+R] \right\}, 
\end{align} 
which is of the exponential order of
\begin{align}
\max_{\rho_{UU'}} M(\rho_{UU'}) \exp\left\{-n [E(\hat{\sigma}_s^2,\rho_{UU'})+R] \right\},
\end{align}
where $M(\rho_{UU'})$ is the typical number of codeword pairs $(\bu_{k},\bu_{k'})$ whose empirical correlation coefficient is about $\rho_{UU'}$ and the sum is over a fine grid in $(-1,1)$. Since there are about $e^{2n(R+I_{US})}$ codeword pairs in $\calC_{n}$ and since the probability of the event $\{\bu_{k}^{T}\bu_{k'}/(nW) \approx \rho_{UU'}\}$ is about $\exp\{-\frac{n}{2} \ln[1/(1-\rho_{UU'}^{2})]\}$, the typical value of the number $M(\rho_{UU'})$ is of the exponential order of $\exp\{-n[2(R+I_{US})-\frac{1}{2} \ln[1/(1-\rho_{UU'}^{2})]]\}$, whenever $2(R+I_{US})-\frac{1}{2} \ln[1/(1-\rho_{UU'}^{2})]>0$, and is zero otherwise. More details on this point can be found in \cite{MERHAV_TYPICAL, MERHAV_GAUSS}.  
One can arrive at the constraint in \eqref{EXexponent} by performing a standard expurgation process (like the one provided in \cite[Appendix A]{TM_UD}) according to the conditional error probabilities.

\section*{Appendix A - Bin Index Decoding}
\renewcommand{\theequation}{A.\arabic{equation}}
\setcounter{equation}{0}

In principle, both in the DPC and the GP channel models defined in Subsections \ref{SEC_Objectives} and \ref{SEC_GP_setup}, respectively, an optimal bin index decoder should be implemented, in order to minimize the probability of error. Such an optimal decoder compares between `metrics' that depend on the whole set of $\bu$'s in each bin. Therefore, this optimal bin index decoder is relatively hard to implement, and moreover, it is quite complicated to analyze. Considering the papers \cite{AM2018} and \cite{MERHAV2014}, where it has been proved that specific sub-optimal bin index decoders attain the same random coding exponent as the optimal bin index decoder, it is reasonable to suspect that also in the current work, one may lose nothing, at least in the random coding exponent sense, when using a two-stage sub-optimal decoder, like the one in \eqref{decodek} and \eqref{decodem}. In the lines to follow, we show by relatively simple arguments that this is indeed the case here.  

Consider a random codebook of size $M_0=e^{nR_0}$, partitioned into $M=e^{n(R_0-R_B)}$ bins, each of size $M_B=e^{nR_B}$.
Let $m\in\{0,1,\ldots,M_0-1\}$ be the index of the transmitted codeword, and let $\mu=\lfloor m/M_B \rfloor\in\{0,1,\ldots,
M-1\}$ be the index of its bin, $\calC_{\mu}$. Consider a sub-optimal, two--step decoder that decodes $m$ by 
first decoding $\mu$, using an optimal bin index decoder, and then decoding $m$ by ML decoding within the sub--code 
pertaining to the decoded bin. Then, on the one hand,
\begin{equation}
\label{ToRef1}
\pr\{\hat{m}\ne m\} \gexe e^{-nE_{\mbox{\tiny r}}(R_0)}.
\end{equation}
On the other hand,
\begin{eqnarray}
\pr\{\hat{m}\ne m\}&=& 
\pr\{\hat{m}\ne m,~\hat{\mu}\ne\mu\}+
\pr\{\hat{m}\ne m,~\hat{\mu}=\mu\} \\
&\le&\pr\{\hat{\mu}\ne\mu\}+
\pr\left\{\mbox{arg max}_{\bu_{m'}\in\calC_\mu} P(\by|\bu_{m'})\ne m,~\hat{\mu}=\mu\right\} \\
&\le&\pr\{\hat{\mu}\ne\mu\}+
\pr\left\{\mbox{arg max}_{\bu_{m'}\in\calC_\mu} P(\by|\bu_{m'})\ne m\right\} \\
\label{ToRef2}
&\le&\pr\{\hat{\mu}\ne\mu\}+e^{-nE_{\mbox{\tiny r}}(R_B)},
\end{eqnarray}
and so, it follows from \eqref{ToRef1} and \eqref{ToRef2} that
\begin{eqnarray}
\pr\{\hat{\mu}\ne\mu\}&\gexe&
e^{-nE_{\mbox{\tiny r}}(R_0)}-e^{-nE_{\mbox{\tiny r}}(R_B)} \\
\label{ToRef3}
&\exe&e^{-nE_{\mbox{\tiny r}}(R_0)}.
\end{eqnarray}
Now, let $\tilde{m}$ be the ordinary ML--decoded version of $m$ and let $\tilde{\mu}=\lfloor \tilde{m}/M_B\rfloor$ be a sub-optimal bin index decoder that relies simply on $\tilde{m}$. 
Then, as a matching upper bound to \eqref{ToRef3} we have
\begin{eqnarray}
\pr\{\hat{\mu}\ne\mu\}&\le&
\pr\{\tilde{\mu}\ne\mu\} \\
&\le&\pr\{\tilde{m}\ne m\} \\
\label{ToRef4}
&\le& e^{-nE_{\mbox{\tiny r}}(R_0)}.
\end{eqnarray}
We conclude that 
\begin{align}
\pr\{\tilde{\mu}\ne\mu\}
\exe \pr\{\hat{\mu}\ne\mu\},
\end{align} 
i.e., at least in the random coding sense, the suggested sub-optimal bin index decoder is as good as the optimal bin index decoder.

\section*{Appendix B - Proofs of Lemmas \ref{lemmaK} and \ref{lemmaRHO}}
\renewcommand{\theequation}{B.\arabic{equation}}
\setcounter{equation}{0}

\subsection*{Proof of Lemma \ref{lemmaK}}

	Consider first a given $m$ and $\bs\in\calS(\sqrt{n\hat{Q}})$.
	Observe that $K_n(\bs,m)=|\calC_m\cap\calT(\bU|\bs)|$ is a
	binomial random variable with $|\calC_m|=M_2=e^{n[I_{US}+\Delta]}$ trials and probability of
	success of the exponential order of $e^{-nI_{US}}$,
	and therefore $K_n(m,s)$
	concentrates double--exponentially rapidly around $e^{n\Delta}$. 
	More precisely, similarly as in \cite[Eqs.\ (24)--(25)]{MERHAV2017}, given $0< \epsilon \ll \Delta$,
	we have:
	\begin{eqnarray}
	\label{Ku}
	\pr\{K_n(\bs,m)> e^{n(\Delta+\epsilon)}\}&\le&\exp\{-(n\epsilon-1)e^{n\epsilon}\}\\
	\label{Kd}
	\pr\{K_n(\bs,m)< e^{n(\Delta-\epsilon)}\}&\le&\exp\{-[1-(n\epsilon+1)e^{-n\epsilon}])e^{n\epsilon}\}.
	\end{eqnarray}
	Let $\calG_1^n$ be the collection of codes for which $e^{n(\Delta-\epsilon)}\le K_n(\bs,m)\le e^{n(\Delta+\epsilon)}$
	for all $m\in\{0,1,\ldots,M-1\}$ and $\bs\in\calS(\sqrt{n\hQ})$. 
	We now show that a randomly selected code falls in $\calG_1^n$
	with a probability that tends to one double--exponentially. To this end,
	consider a fine quantization of the vectors in $\calS_n(\sqrt{n\hQ})$ in the following manner: 
	every $\bs\in\calS_n(\sqrt{n\hQ})$
	can be represented as follows. Let $\theta_i\in[0,2\pi)$, $i=1,2,\ldots,n-1$, 
	and let the components of $\bs$ be given by
	\begin{eqnarray}
	s_1&=&\sqrt{nQ}\sin\theta_1,\\
	s_2&=&\sqrt{nQ}\cos\theta_1\sin\theta_2,\\
	s_3&=&\sqrt{nQ}\cos\theta_1\cos\theta_2\sin\theta_3,\\
	&\ldots&\nonumber\\
	s_{n-1}&=&\sqrt{nQ}\cos\theta_1\cdot\cdot\cdot\cos\theta_{n-2}\sin\theta_{n-1},\\
	s_{n}&=&\sqrt{nQ}\cos\theta_1\cdot\cdot\cdot\cos\theta_{n-2}\cos\theta_{n-1}.
	\end{eqnarray}
	Now, let us quantize each $\theta_i$ uniformly using $n^2$ points with a quantization step-size of $2\pi/n^2$.
	Thus, the total number of quantization points within the hyprcube $[0,2\pi)^{n-1}$ is $n^{2(n-1)}$.
	Given $\bs\in\calS_n(\sqrt{n\hQ})$, we extract $\theta_1,\ldots\theta_{n-1}$, and then quantize each
	$\theta_i$ to its nearest quantization value, $\hat{\theta}_i$. Note that $|\sin\hat{\theta}_i-\sin\theta_i|\le
	|\hat{\theta}_i-\theta_i|\le\pi/n^2$ since the absolute value of the derivative of the sinusoidal function is bounded by 1.
	The same comment applies also to $|\cos\hat{\theta}_i-\cos\theta_i|$.
	According to the above representation, we can think of each component $s_k$, $k=1,2,\ldots,n$, as 
	being given by $\sqrt{n\hQ}\prod_{i=1}^ng_{i,k}(\theta_i)$, where
	each $g_{i,k}(\theta_i)$ is either $\sin\theta_i$, or $\cos\theta_i$, or it is identical to $1$.
	Clearly, it is also true that $|g_i(\hat{\theta}_i)-g_i(\theta_i)|\le|\hat{\theta}_i-\theta_i|\le\pi/n^2$.
	and that $|g_i(\theta_i)|$ and $|g_i(\hat{\theta}_i)|$ are upper bounded by 1.
	The quantization error in $s_\ell$, $\ell=1,2,\ldots,n$, is therefore upper bounded by
	\begin{eqnarray}
	|s_\ell-\hat{s}_\ell|&=&\sqrt{n\hQ}\bigg|\prod_{i=1}^ng_i(\theta_i)-\prod_{i=1}^ng_i(\hat{\theta}_i)\bigg|\\
	&=&\sqrt{n\hQ}\Bigg|\sum_{j=1}^{n}\left[\prod_{i=1}^{j-1}g_i(\hat{\theta}_i)\prod_{i=j}^ng_i(\theta_i)-
	\prod_{i=1}^{j}g_i(\hat{\theta}_i)\prod_{i=j+1}^ng_i(\theta_i)\right]\Bigg|\\
	&\le&\sqrt{n\hQ}\sum_{j=1}^{n}\bigg|\prod_{i=1}^{j-1}g_i(\hat{\theta}_i)\prod_{i=j}^ng_i(\theta_i)-
	\prod_{i=1}^{j}g_i(\hat{\theta}_i)\prod_{i=j+1}^ng_i(\theta_i)\bigg|\\
	&\le&\sqrt{n\hQ}\sum_{j=1}^{n}\Bigg|\prod_{i=1}^{j-1}g_i(\hat{\theta}_i)\prod_{i=j+1}^ng_i(\theta_i)
	\cdot[g_j(\hat{\theta_j})-g_j(\theta_j)]\Bigg|\\
	&\le&\sqrt{n\hQ}\sum_{j=1}^{n}\Bigg|\prod_{i=1}^{j-1}g_i(\hat{\theta}_i)
	\prod_{i=j+1}^ng_i(\theta_i)\Bigg|\cdot
	\bigg|g_j(\hat{\theta}_j)-g_j(\theta_j)\bigg|\\
	&\le&\sqrt{n\hQ}\sum_{j=1}^{n}\frac{\pi}{n^2}\\
	&=&\frac{\pi\sqrt{\hQ}}{\sqrt{n}},
	\end{eqnarray}
	and so, $\|\bs-\hat{\bs}\|\le\pi\sqrt{\hQ}$.
	Since the large--deviations bounds on $K_n(\bs,m)$ (eqs.\ (\ref{Ku}) and (\ref{Kd})), 
	decay double--exponentially with $n$, while $M$ is only exponential and
	the number of quantization points is only $n^{2(n-1)}$, it follows by the union bound 
	that with probability that tends to one double-exponentially,
	these bounds hold simulatenously for all $m$ and all quantization points on the hypersphere surface.
	Given that this happens, then for any point on the hypersphere surface, we have
	\begin{eqnarray}
	\bu^T\bs&=&\bu^T[\hat{\bs}+(\bs-\hat{\bs})]\\
	&=&\bu^T\hat{\bs}+\bu^T(\bs-\hat{\bs})\\
	&\ge&\alpha_{\bs}\hat{\bs}^T\hat{\bs}-\|\bu\|\cdot\|\bs-\hat{\bs}\|\\
	&\ge&\alpha_{\bs}\hat{\bs}^T\hat{\bs}-\sqrt{nW}\cdot\pi\sqrt{\hQ}\\
	&=&n\alpha\sqrt{Q\hat{Q}}-\pi\sqrt{nW\hQ},
	\end{eqnarray}
	where the second term, which scales with $\sqrt{n}$, is asymptotically 
	negligible relative to the first one, which is linear in $n$.
	It follows that every codeword in $\calC_m\cap\calT(\bU|\hat{\bs})$ is also in
	$\calC_m\cap\calT(\bU|\bs)$, provided that $\bs$ is quantized to $\hat{\bs}$, and provided that 
	$\calT(\bU|\bs)$ is slightly expanded by reducing the 
	threshold, $\alpha_{\bs}\|\bs\|^2$, within a relative amount that vanishes as $n$ grows without bound.

\subsection*{Proof of Lemma \ref{lemmaRHO}}

	For a given $\by$ and $\tilde{k}$, consider first the probability
	$$\pr\left\{\hat{\rho}^2(\bU_{\tilde{k}},\by)\ge 1-\exp[-2(R+I_{US}-\epsilon)]\right\}.$$
	Since $\bU_{\tilde{k}}$ is distributed uniformly over the surface of a hypersphere,
	this probability is equal to the relative area of an $n$-dimensional spherical cap with half-angle
	\begin{align}
	\theta=\arccos(\sqrt{1-\exp[-2(R+I_{US}-\epsilon)]}),
	\end{align} 
	which is of the exponential order of
	(see, e.g., \cite{WYNER}):
	\begin{eqnarray}
	\sin^n(\theta)&=&\exp\left\{\frac{n}{2}\ln\sin^2\theta\right\}\nonumber\\
	&=&\exp\left\{\frac{n}{2}\ln\left(1-\cos^2\theta\right)\right\}\nonumber\\
	&=&\exp\left\{\frac{n}{2}\ln\left(1-\left[1-\exp\{-2(R+I_{US}-\epsilon\}\right]\right)\right\}\nonumber\\
	&=&\exp\{-n(R+I_{US}-\epsilon)\}.
	\end{eqnarray}
	Therefore, for a given $(k,k')$ and $\by$,
	\begin{align}
	&\pr\left\{\max_{\tilde{k}\ne k,k'}\hat{\rho}^2(\bU_{\tilde{k}},\by)< 
	1-\exp[-2(R+I_{US}-\epsilon)]\right\}\nonumber\\
	&~~~~\exe\left[1-\exp\{-n(R+I_{US}-\epsilon)\}\right]^{M}\nonumber\\
	&~~~~=\left[1-\exp\{-n(R+I_{US}-\epsilon)\}\right]^{\exp\{n[R+I_{US}+\Delta]\}}\nonumber\\
	&~~~~=\exp\left\{\exp\{n[R+I_{US}+\Delta]\}\ln\left[1-\exp\{-n(R+I_{US}-\epsilon)\}\right]\right\}\nonumber\\
	&~~~~\le\exp\left\{-\exp\{n[R+I_{US}+\Delta]\}\exp\{-n(R+I_{US}-\epsilon)\}\right\}\nonumber\\
	&~~~~=\exp\left\{-\exp\{n[\Delta+\epsilon]\}\right\},
	\end{align}
	which decays double-exponentially as $n\to\infty$.
	Applying the union bound over all pairs $(k,k')$, introduces an exponential factor of the order of $M^2$, which
	leaves the probability of the union double-exponentially small.
	As for the union over $\{\by\}$, 
	it is sufficient to consider one hypersphere surface since the empirical correlation coefficient
	is invariant to scaling of $\by$. 
	Consider again, quantization of the hypersphere of $\by$ as in the proof of Lemma \ref{lemmaK}, with $n^{2(n-1)}$
	quantization points. 
	With very high probability, the same property applies to every quantization point at the same time, since
	the factor of $n^{2(n-1)}$ still leaves the probability of the union double-exponentially small.
	Finally, we pass to the continuum of $\{\by\}$ in the hypersphere, 
	in the same way as in the proof of Lemma \ref{lemmaK}, 
	by using the fact that the quantization error is small and hence 
	affects the empirical correlation coefficient by a vanishingly small amount as $n$ grows without bound.

\section*{Appendix C - Proof of Theorem \ref{THEOREM1}}
\renewcommand{\theequation}{C.\arabic{equation}}
\setcounter{equation}{0}

\subsection*{Step 1: Error exponent for a general decoding metric}
Given $m \in \{1,2,\ldots,M\}$, $\bS=\bs$, $\bU=\bu \in \calC(m,\bs)$, $\bX=\bx \in \calT(Q_{X|SU}|\bs,\bu)$, and $\bY=\by$, the probability of error is given by
\begin{align}
P_{\mbox{\tiny e}}(m,\bs,\bu,\bx,\by) 
&= \pr \left\{\bigcup_{Q_{S'}} \bigcup_{m' \neq m} \bigcup_{k=1}^{M(Q_{S'})} \{G(\hat{P}_{\bU_{Q_{S'},m',k}\by},Q_{S'}) \geq G(\hat{P}_{\bu\by},\hat{P}_{\bs})\} \right\} \\
\label{ref0}
&\doteq \min \left\{1, \sum_{Q_{S'}} \sum_{m' \neq m} \sum_{k=1}^{M(Q_{S'})} \pr \{G(\hat{P}_{\bU_{Q_{S'},m',k}\by},Q_{S'}) \geq G(\hat{P}_{\bu\by},\hat{P}_{\bs})\} \right\}.
\end{align}
For any $Q_{S'}$, $m' \neq m$, and $k \in \{1,2,\ldots,M(Q_{S'})\}$, let us denote $\bU' = \bU_{Q_{S'},m',k}$. Then, the pairwise probability inside \eqref{ref0} can be assessed using the method of types as follows 
\begin{align}
&\pr \{G(\hat{P}_{\bU'\by},Q_{S'}) \geq G(\hat{P}_{\bu\by},\hat{P}_{\bs})\} \nn \\
&~~~= \sum_{\{\bu':~ G(\hat{P}_{\bu'\by},Q_{S'}) \geq G(\hat{P}_{\bu\by},\hat{P}_{\bs})\}} \frac{1}{|\calT(Q_{U'})|} \\
&~~~= \sum_{\{Q_{U'|Y}:~ G(Q_{U'|Y} \times \hat{P}_{\by},Q_{S'}) \geq G(\hat{P}_{\bu\by},\hat{P}_{\bs})\}} \sum_{\bu' \in \calT(Q_{U'|Y}|\by)} \frac{1}{|\calT(Q_{U'})|} \\
&~~~= \sum_{\{Q_{U'|Y}:~ G(Q_{U'|Y} \times \hat{P}_{\by},Q_{S'}) \geq G(\hat{P}_{\bu\by},\hat{P}_{\bs})\}} \frac{|\calT(Q_{U'|Y}|\by)|}{|\calT(Q_{U'})|} \\
&~~~\doteq \sum_{\{Q_{U'|Y}:~ G(Q_{U'|Y} \times \hat{P}_{\by},Q_{S'}) \geq G(\hat{P}_{\bu\by},\hat{P}_{\bs})\}} \exp\{-n I_{Q}(U';Y)\} \\
\label{ref1}
&~~~\doteq \exp\left\{-n \min_{\{Q_{U'|Y}:~ G(Q_{U'|Y} \times \hat{P}_{\by},Q_{S'}) \geq G(\hat{P}_{\bu\by},\hat{P}_{\bs})\}} I_{Q}(U';Y)\right\}.
\end{align}
Note that the expression in \eqref{ref1} does not depend on $m' \neq m$ and $k \in \{1,2,\ldots,M(Q_{S'})\}$, hence, substituting \eqref{ref1} back into \eqref{ref0} yields that
\begin{align}
&P_{\mbox{\tiny e}}(m,\bs,\bu,\bx,\by) \nn \\
&~~\doteq \min \left[ 1, \sum_{Q_{S'}} \sum_{m' \neq m} \sum_{k=1}^{M(Q_{S'})} \exp\left\{-n \min_{\{Q_{U'|Y}:~ G(Q_{U'|Y} \times \hat{P}_{\by},Q_{S'}) \geq G(\hat{P}_{\bu\by},\hat{P}_{\bs})\}} I_{Q}(U';Y)\right\} \right] \\
&~~\doteq \exp\left\{-n \min_{\{Q_{S'},Q_{U'|Y}:~ G(Q_{U'|Y} \times \hat{P}_{\by},Q_{S'}) \geq G(\hat{P}_{\bu\by},\hat{P}_{\bs})\}} [I_{Q}(U';Y) - R(Q_{S'})-R-\eps]_{+}\right\} \\
&~~\dfn \exp \left\{-n E_{0}(R,\hat{P}_{\bs},\hat{P}_{\bu\by}) \right\}.
\end{align}
Next, given $m \in \{1,2,\ldots,M\}$, $\bS=\bs$, $\bU=\bu \in \calC(m,\bs)$, and $\bX=\bx \in \calT(Q_{X|SU}|\bs,\bu)$, we average over the randomness of the channel output vector $\bY$:
\begin{align}
P_{\mbox{\tiny e}}(m,\bs,\bu,\bx) 
&\doteq \sum_{\by \in \calY^{n}} W(\by|\bx,\bs) \exp \left\{-n E_{0}(R,\hat{P}_{\bs},\hat{P}_{\bu\by}) \right\} \\  
&= \sum_{Q_{Y|SUX}}\sum_{\by \in \calT(Q_{Y|SUX}|\bs,\bu,\bx)} W(\by|\bx,\bs) \exp \left\{-n E_{0}(R,\hat{P}_{\bs},\hat{P}_{\bu\by}) \right\} \\
&\doteq \sum_{Q_{Y|SUX}} \exp\{-n D(Q_{Y|SUX} \| W_{Y|XS}|\hat{P}_{\bs \bu \bx})\} \exp \left\{-n E_{0}(R,\hat{P}_{\bs},Q_{UY}) \right\} \\
\label{ref2}
&\doteq \exp\left\{-n \cdot \min_{Q_{Y|SUX}} \{ D(Q_{Y|SUX} \| W_{Y|XS}|\hat{P}_{\bs \bu \bx}) + E_{0}(R,\hat{P}_{\bs},Q_{UY}) \} \right\}.
\end{align}
Note that \eqref{ref2} depends on $\bs,\bu,\bx$ only via their joint empirical distribution, hence, averaging over the randomness of $\bU \in \calB(\hat{P}_{\bs},m) \cap \calT(Q_{U|S}|\bs)$ and $\bX \in \calT(Q_{X|SU}|\bs,\bu)$ yields
\begin{align}
P_{\mbox{\tiny e}}(m,\bs) 
&\doteq \exp\left\{-n \cdot \min_{Q_{Y|SUX}} \{ D(Q_{Y|SUX} \| W_{Y|XS}|Q_{UX|S} \times \hat{P}_{\bs}) + E_{0}(R,\hat{P}_{\bs},Q_{UY}) \} \right\}  \\
\label{ref5}
&\dfn \exp \{-n \cdot E_{1}(R,\hat{P}_{\bs}, Q_{UX|S})\}.
\end{align}
The exponent function in \eqref{ref5} depends on $Q_{UX|S}$, which are design parameters of the system that can be chosen to minimize the average error probability. Thus, minimizing the average probability of error w.r.t.\ $Q_{UX|S}$, we arrive at
\begin{align}
P_{\mbox{\tiny e}}(m,\bs) 
&\doteq \exp \left\{-n \cdot \max_{Q_{UX|S}} E_{1}(R,\hat{P}_{\bs}, Q_{UX|S})\right\} \\
&\dfn \exp \{-n \cdot E_{2}(R,\hat{P}_{\bs})\}.
\end{align}
Averaging w.r.t.\ the randomness of $\bS$, we arrive at 
\begin{align}
P_{\mbox{\tiny e}}(m)
&\doteq \sum_{\bs \in \calS^{n}} P(\bs) \exp \{-n \cdot E_{2}(R,\hat{P}_{\bs})\} \\
&= \sum_{Q_{S}} \sum_{\bs \in \calT(Q_{S})} P(\bs) \exp \{-n \cdot E_{2}(R,\hat{P}_{\bs})\} \\
&\doteq \exp\left\{-n \cdot \min_{Q_{S}} [D(Q_{S}\|P_{S}) + E_{2}(R,Q_{S})] \right\},
\end{align}
which does not depend on the specific value of $m$, hence,
\begin{align}
P_{\mbox{\tiny e}}
\doteq \exp\left\{-n \cdot \min_{Q_{S}} [D(Q_{S}\|P_{S}) + E_{2}(R,Q_{S})] \right\}.
\end{align}
Finally, due to the arbitrariness of $\eps > 0$, it follows that the exact random coding error exponent is given by
\begin{align}
E(R) = \min_{Q_{S}} \max_{Q_{UX|S}} [D(Q_{S}\|P_{S}) + E_{1}(R,Q_{S},Q_{UX|S})],
\end{align} 
where 
\begin{align}
E_{1}(R,Q_{S},Q_{UX|S}) = 
\min_{Q_{Y|SUX}} \{ D(Q_{Y|SUX} \| W_{Y|XS}|Q_{UX|S} \times Q_{S}) + E_{0}(R,Q_{S},Q_{UY}) \},
\end{align}
and 
\begin{align}
\label{ref3}
E_{0}(R,Q_{S},Q_{UY}) = 
\min_{\{Q_{S'},Q_{U'|Y}:~ G(Q_{U'Y},Q_{S'}) \geq G(Q_{UY},Q_{S})\}} [I_{Q}(U';Y) - R(Q_{S'}) - R]_{+}.
\end{align}

\subsection*{Step 2: An upper bound on the error exponent}

Note that the minimum in \eqref{ref3} can be upper-bounded by choosing specific distributions in the feasible set. In \eqref{ref3}, we take $Q_{U'|Y}=Q_{U|Y}$, $Q_{S'}=Q_{S}$, and then  
\begin{align}
E_{0}(R,Q_{S},Q_{UY}) \leq [I_{Q}(U;Y) - R(Q_{S}) - R]_{+}.
\end{align}
Hence, the overall error exponent is upper-bounded as 
\begin{align}
\label{ref4}
E(R) 
\leq \min_{Q_{S}} \max_{Q_{UX|S}} \min_{Q_{Y|SUX}} \{&D(Q_{S}\|P_{S}) + D(Q_{Y|SUX} \| W_{Y|XS}|Q_{UX|S} \times Q_{S}) \nn \\
&~+ [I_{Q}(U;Y) - R(Q_{S}) - R]_{+} \}.
\end{align} 

\subsection*{Step 3: An optimal universal decoder}

We prove that the upper bound of \eqref{ref4} is attainable by choosing the universal decoding metric $G(Q_{UY},Q_{S}) = I_{Q}(U;Y) - R(Q_{S})$. Now, we get for \eqref{ref3}
\begin{align}
E_{0}(R,Q_{S},Q_{UY}) 
&= \min_{\{Q_{S'},Q_{U'|Y}:~ G(Q_{U'Y},Q_{S'}) \geq G(Q_{UY},Q_{S})\}} \left[I_{Q}(U';Y) - R(Q_{S'})-R\right]_{+} \\
&= \min_{\left\{\substack{Q_{S'},Q_{U'|Y}: \\ 
I_{Q}(U';Y) - R(Q_{S'}) \geq I_{Q}(U;Y) - R(Q_{S})} \right\}} \left[I_{Q}(U';Y) - R(Q_{S'}) -R \right]_{+} \\
&= \left[I_{Q}(U;Y) - R(Q_{S}) - R\right]_{+},
\end{align}
which completes the proof of Theorem \ref{THEOREM1}.

\section*{Appendix D - Proof of Theorem \ref{THEOREM_TRC}}
\renewcommand{\theequation}{D.\arabic{equation}}
\setcounter{equation}{0}

Assume that the matrix $\{Q_{UX|S}\}$ is fixed.
Let us choose the universal decoding metric $g(Q_{UY},Q_{S}) = I_{Q}(U;Y) - R(Q_{S})$.
For a given codebook, the probability of error is given by
\begin{align}
P_{\mbox{\tiny e}}(\calC_{n})
\label{REF0}
&= \sum_{Q_{S}} \pr\left[\bS \in \calT(Q_{S})\right] \frac{1}{M\cdot M(Q_{S})} \sum_{m=1}^{M} \sum_{\ell=1}^{M(Q_{S})} \sum_{\bs \in \calT(Q_{S})} P(\bs|\bu_{m,\ell}) \nn \\
&~~\times \sum_{\bx \in \calT(Q_{X|US}|\bu_{m,\ell},\bs)} \frac{1}{|\calT(Q_{X|US}|\bu_{m,\ell},\bs)|} 
 \nn \\
&~~\times \sum_{\by \in \calY^{n}} W(\by|\bx,\bs) \IND\left\{\bigcup_{Q_{S'}} \bigcup_{m' \neq m} \bigcup_{\ell'=1}^{M(Q_{S'})} \{g(\hat{P}_{\bu_{m',\ell'}\by},Q_{S'}) \geq g(\hat{P}_{\bu_{m,\ell}\by},Q_{S})\} \right\}.
\end{align}
We upper-bound the inner-most summation over $\by$ by\footnote{A tighter upper bound in \eqref{Ineqality_to_improve} can be obtained by using the inequality \cite{GAL68}
\begin{align} \label{Improved}
\IND\left\{\bigcup_{k} 
\{\exp\{ng(\hat{P}_{\bu_{k}\by})\} \geq \exp\{ng(\hat{P}_{\bu\by})\}\} \right\}
\leq \left(\sum_{k} \inf_{\mu \geq 0} \left[\frac{\exp\{ng(\hat{P}_{\bu_{k}\by})\}}{\exp\{ng(\hat{P}_{\bu\by})\}}\right]^{\mu} \right)^{\rho},
\end{align}
where $\rho >0$ is a parameter to be optimized. Using this inequality is expected to yield an exponent function which will be tighter at relatively high coding rates. Since the random coding error exponent given in Theorem \ref{THEOREM1} provides the true exponential behavior at the range of high coding rates, we refrain from applying \eqref{Improved} for two reasons: (i) simplicity and (ii) at low rates, the optimal value of $\rho$ is expected to be 1 anyway (after limiting $\rho$ to be less than 1 in order to apply Jensen's inequality), which corresponds to a simple union bound.}:
\begin{align}
&\sum_{\by \in \calY^{n}} W(\by|\bx,\bs) \IND\left\{\bigcup_{Q_{S'}} \bigcup_{m' \neq m} \bigcup_{\ell'=1}^{M(Q_{S'})} \{g(\hat{P}_{\bu_{m',\ell'}\by},Q_{S'}) \geq g(\hat{P}_{\bu_{m,\ell}\by},Q_{S})\} \right\} \nn \\
&~~=\sum_{\by \in \calY^{n}} W(\by|\bx,\bs) \IND\left\{\bigcup_{Q_{S'}} \bigcup_{m' \neq m} \bigcup_{\ell'=1}^{M(Q_{S'})} \{\exp\{ng(\hat{P}_{\bu_{m',\ell'}\by},Q_{S'})\} \geq \exp\{ng(\hat{P}_{\bu_{m,\ell}\by},Q_{S})\} \} \right\}  \nn \\
\label{Ineqality_to_improve}
&~~\leq \sum_{\by \in \calY^{n}} W(\by|\bx,\bs) 
\sum_{Q_{S'}} 
\sum_{m' \neq m} 
\sum_{\ell'=1}^{M(Q_{S'})}
\inf_{\mu \geq 0} \left[\frac{\exp\{ng(\hat{P}_{\bu_{m',\ell'}\by},Q_{S'})\}}{\exp\{ng(\hat{P}_{\bu_{m,\ell}\by},Q_{S})\}}\right]^{\mu} \\
&~~=
\sum_{Q_{S'}} 
\sum_{m' \neq m} 
\sum_{\ell'=1}^{M(Q_{S'})}
\sum_{\by \in \calY^{n}} W(\by|\bx,\bs)
\inf_{\mu \geq 0} \left[\frac{\exp\{ng(\hat{P}_{\bu_{m',\ell'}\by},Q_{S'})\}}{\exp\{ng(\hat{P}_{\bu_{m,\ell}\by},Q_{S})\}}\right]^{\mu},
\end{align}
where the inner-most sum over $\by \in \calY^{n}$ is assessed by the method of types as
\begin{align}
&\sum_{\by \in \calY^{n}} W(\by|\bx,\bs)
\exp\left\{-n \cdot \sup_{\mu \geq 0} \left\{ \mu \cdot [g(\hat{P}_{\bu_{m,\ell}\by},Q_{S}) - g(\hat{P}_{\bu_{m',\ell'}\by},Q_{S'})] \right\} \right\} \\
&=  
\sum_{Q_{Y|UU'SX}}
\sum_{\by \in \calT(Q_{Y|UU'SX}|\bu_{m,\ell},\bu_{m',\ell'},\bs,\bx)} W(\by|\bx,\bs)
\exp\left\{-n \cdot \Psi(\hat{P}_{\bu_{m,\ell}\by},Q_{S},\hat{P}_{\bu_{m',\ell'}\by},Q_{S'})  \right\} \\
&= 
\sum_{Q_{Y|UU'SX}}
\sum_{\by \in \calT(Q_{Y|UU'SX}|\bu_{m,\ell},\bu_{m',\ell'},\bs,\bx)} 
e^{n \mathbb{E}_{Q} [\ln W(Y|X,S)]}
\exp\left\{-n \cdot \Psi(Q_{UY},Q_{S},Q_{U'Y},Q_{S'})\right\} \\
&\doteq   
\sum_{Q_{Y|UU'SX}}
e^{-n D(Q_{Y|UU'SX}\|W_{Y|SX}|\hat{P}_{\bu_{m,\ell},\bu_{m',\ell'},\bs,\bx})}
\exp\left\{-n \cdot \Psi(Q_{UY},Q_{S},Q_{U'Y},Q_{S'})\right\} \\
&\DEF \exp\{-n E_{0}(\hat{P}_{\bu_{m,\ell},\bu_{m',\ell'},\bs,\bx},Q_{S'})\}.
\end{align}
Substituting it back into \eqref{REF0} yields that
\begin{align}
&P_{\mbox{\tiny e}}(\calC_{n}) \nn \\
&\lexe \sum_{Q_{S}} \pr\left[\bS \in \calT(Q_{S})\right] \frac{1}{M\cdot M(Q_{S})} \sum_{m=1}^{M} \sum_{\ell=1}^{M(Q_{S})} \sum_{\bs \in \calT(Q_{S})} P(\bs|\bu_{m,\ell}) \nn \\
&~~\times \sum_{\bx \in \calT(Q_{X|US}|\bu_{m,\ell},\bs)} \frac{1}{|\calT(Q_{X|US}|\bu_{m,\ell},\bs)|} 
\sum_{Q_{S'}} 
\sum_{m' \neq m} 
\sum_{\ell'=1}^{M(Q_{S'})} \exp\{-n E_{0}(\hat{P}_{\bu_{m,\ell},\bu_{m',\ell'},\bs,\bx},Q_{S'})\} \\
\label{REFa1}
&= \sum_{Q_{S}} \pr\left[\bS \in \calT(Q_{S})\right] \frac{1}{M\cdot M(Q_{S})} \sum_{m=1}^{M} \sum_{\ell=1}^{M(Q_{S})} \sum_{\bs \in \calT(Q_{S})} P(\bs|\bu_{m,\ell}) \nn \\
&~~\times  
\sum_{Q_{S'}} 
\sum_{m' \neq m} 
\sum_{\ell'=1}^{M(Q_{S'})} 
\sum_{\bx \in \calT(Q_{X|US}|\bu_{m,\ell},\bs)} \frac{1}{|\calT(Q_{X|US}|\bu_{m,\ell},\bs)|}
\exp\{-n E_{0}(\hat{P}_{\bu_{m,\ell},\bu_{m',\ell'},\bs,\bx},Q_{S'})\}.
\end{align}
The inner-most sum over $\bx$ is assessed by the method of types as
\begin{align}
&\sum_{\bx \in \calT(Q_{X|US}|\bu_{m,\ell},\bs)} \frac{1}{|\calT(Q_{X|US}|\bu_{m,\ell},\bs)|}
\exp\{-n E_{0}(\hat{P}_{\bu_{m,\ell},\bu_{m',\ell'},\bs,\bx},Q_{S'})\} \nn \\
&= \sum_{\{\bar{Q}_{X|UU'S}:~\bar{Q}_{X|US}=Q_{X|US}\}}
\sum_{\bx \in \calT(\bar{Q}_{X|UU'S}|\bu_{m,\ell},\bu_{m',\ell'},\bs)} \frac{1}{|\calT(Q_{X|US}|\bu_{m,\ell},\bs)|}
\exp\{-n E_{0}(\hat{P}_{\bu_{m,\ell},\bu_{m',\ell'},\bs,\bx},Q_{S'})\}  \\
&= \sum_{\{\bar{Q}_{X|UU'S}:~\bar{Q}_{X|US}=Q_{X|US}\}}
 \frac{|\calT(\bar{Q}_{X|UU'S}|\bu_{m,\ell},\bu_{m',\ell'},\bs)|}{|\calT(Q_{X|US}|\bu_{m,\ell},\bs)|}
\exp\{-n E_{0}(\hat{P}_{\bu_{m,\ell},\bu_{m',\ell'},\bs} \times \bar{Q}_{X|UU'S},Q_{S'})\}  \\
&\doteq \sum_{\{\bar{Q}_{X|UU'S}:~\bar{Q}_{X|US}=Q_{X|US}\}}
\exp\{-n I_{\bar{Q}}(U';X|US)\}
\exp\{-n E_{0}(\hat{P}_{\bu_{m,\ell},\bu_{m',\ell'},\bs} \times \bar{Q}_{X|UU'S},Q_{S'})\}  \\
&\doteq 
\exp\left\{-n \cdot \min_{\{\bar{Q}_{X|UU'S}:~\bar{Q}_{X|US}=Q_{X|US}\}} [I_{\bar{Q}}(U';X|US)
+ E_{0}(\hat{P}_{\bu_{m,\ell},\bu_{m',\ell'},\bs} \times \bar{Q}_{X|UU'S},Q_{S'})] \right\}  \\
&\DEF \exp\{-n E_{1}(\hat{P}_{\bu_{m,\ell},\bu_{m',\ell'},\bs},Q_{S'})\}.
\end{align}
Substituting it back into \eqref{REFa1} yields that
\begin{align}
P_{\mbox{\tiny e}}(\calC_{n})
&\lexe \sum_{Q_{S}} \pr\left[\bS \in \calT(Q_{S})\right] \frac{1}{M\cdot M(Q_{S})} \sum_{m=1}^{M} \sum_{\ell=1}^{M(Q_{S})} \sum_{\bs \in \calT(Q_{S})} P(\bs|\bu_{m,\ell}) \nn \\
&~~\times  
\sum_{Q_{S'}} 
\sum_{m' \neq m} 
\sum_{\ell'=1}^{M(Q_{S'})} \exp\{-n E_{1}(\hat{P}_{\bu_{m,\ell},\bu_{m',\ell'},\bs},Q_{S'})\} \\
\label{REFa2}
&= \sum_{Q_{S}} \pr\left[\bS \in \calT(Q_{S})\right] \frac{1}{M\cdot M(Q_{S})} \sum_{m=1}^{M} \sum_{\ell=1}^{M(Q_{S})}  \nn \\
&~~\times  
\sum_{Q_{S'}} 
\sum_{m' \neq m} 
\sum_{\ell'=1}^{M(Q_{S'})} 
\sum_{\bs \in \calT(Q_{S})} P(\bs|\bu_{m,\ell})
\exp\{-n E_{1}(\hat{P}_{\bu_{m,\ell},\bu_{m',\ell'},\bs},Q_{S'})\}.
\end{align}
Again, the inner-most sum is assessed by the method of types as
\begin{align}
&\sum_{\bs \in \calT(Q_{S})} P(\bs|\bu_{m,\ell})
\exp\{-n E_{1}(\hat{P}_{\bu_{m,\ell},\bu_{m',\ell'},\bs},Q_{S'})\}\nn \\
&= \sum_{\{\tilde{Q}_{S|UU'}:~\tilde{Q}_{S}=Q_{S}\}}
\sum_{\bs \in \calT(\tilde{Q}_{S|UU'}|\bu_{m,\ell},\bu_{m',\ell'})} P(\bs|\bu_{m,\ell})
\exp\{-n E_{1}(\hat{P}_{\bu_{m,\ell},\bu_{m',\ell'},\bs},Q_{S'})\} \\
&= \sum_{\{\tilde{Q}_{S|UU'}:~\tilde{Q}_{S}=Q_{S}\}}
\sum_{\bs \in \calT(\tilde{Q}_{S|UU'}|\bu_{m,\ell},\bu_{m',\ell'})} 
\frac{1}{|\calT(\tilde{Q}_{S|U}|\bu_{m,\ell})|}
\exp\{-n E_{1}(\hat{P}_{\bu_{m,\ell},\bu_{m',\ell'}} \times \tilde{Q}_{S|UU'},Q_{S'})\} \\
&= \sum_{\{\tilde{Q}_{S|UU'}:~\tilde{Q}_{S}=Q_{S}\}} 
\frac{|\calT(\tilde{Q}_{S|UU'}|\bu_{m,\ell},\bu_{m',\ell'})|}{|\calT(\tilde{Q}_{S|U}|\bu_{m,\ell})|}
\exp\{-n E_{1}(\hat{P}_{\bu_{m,\ell},\bu_{m',\ell'}} \times \tilde{Q}_{S|UU'},Q_{S'})\} \\
&\doteq \sum_{\{\tilde{Q}_{S|UU'}:~\tilde{Q}_{S}=Q_{S}\}} 
\exp\{-n I_{\tilde{Q}}(S;U'|U)\}
\exp\{-n E_{1}(\hat{P}_{\bu_{m,\ell},\bu_{m',\ell'}} \times \tilde{Q}_{S|UU'},Q_{S'})\} \\
&\doteq  
\exp\left\{-n \cdot \min_{\{\tilde{Q}_{S|UU'}:~\tilde{Q}_{S}=Q_{S}\}} [I_{\tilde{Q}}(S;U'|U) + E_{1}(\hat{P}_{\bu_{m,\ell},\bu_{m',\ell'}} \times \tilde{Q}_{S|UU'},Q_{S'})]\right\} \\
&\DEF \exp\{-n E_{2}(\hat{P}_{\bu_{m,\ell},\bu_{m',\ell'}},Q_{S'})\}.
\end{align}
Substituting it back into \eqref{REFa2} yields that
\begin{align}
&P_{\mbox{\tiny e}}(\calC_{n}) \nn \\
\label{REFb2}
&\lexe \sum_{Q_{S}} \sum_{Q_{S'}} \pr\left[\bS \in \calT(Q_{S})\right] \frac{1}{M\cdot M(Q_{S})} \sum_{m=1}^{M} \sum_{\ell=1}^{M(Q_{S})}   
\sum_{m' \neq m} 
\sum_{\ell'=1}^{M(Q_{S'})} 
\exp\{-n E_{2}(\hat{P}_{\bu_{m,\ell},\bu_{m',\ell'}},Q_{S'})\} \\
&\doteq \sum_{Q_{S}} \sum_{Q_{S'}} e^{-n D(Q_{S}\|P_{S})} e^{-n (R + R(Q_{S}))} \sum_{m=1}^{M} \sum_{\ell=1}^{M(Q_{S})}   
\sum_{m' \neq m} 
\sum_{\ell'=1}^{M(Q_{S'})} 
\exp\{-n E_{2}(\hat{P}_{\bu_{m,\ell},\bu_{m',\ell'}},Q_{S'})\} \\
&= \sum_{Q_{S}} \sum_{Q_{S'}} e^{-n D(Q_{S}\|P_{S})} e^{-n (R + R(Q_{S}))} 
\sum_{\ddot{Q}_{UU'} \in \calQ(Q_{U},Q_{U'})} N(\ddot{Q}_{UU'})
\exp\{-n E_{2}(\ddot{Q}_{UU'},Q_{S'})\},
\end{align}
where $\calQ(Q_{U},Q_{U'})=\{\ddot{Q}_{UU'}:~\ddot{Q}_{U}=Q_{U},\ddot{Q}_{U'}=Q_{U'}\}$ and
\begin{align}
N(\ddot{Q}_{UU'})
\dfn 
\sum_{m=1}^{M} \sum_{\ell=1}^{M(Q_{S})}   
\sum_{m' \neq m} 
\sum_{\ell'=1}^{M(Q_{S'})}
\IND \left\{ (\bu_{m,\ell},\bu_{m',\ell'}) \in \calT(\ddot{Q}_{UU'}) \right\}.
\end{align}
Now, for any $\rho > 1$,
\begin{align}
\mathbb{E} \left[P_{\mbox{\tiny e}}(\calC_{n})^{1/\rho}\right]
\label{REFa3}
&\lexe \sum_{Q_{S}} \sum_{Q_{S'}} e^{-n D(Q_{S}\|P_{S})/\rho} e^{-n (R + R(Q_{S}))/\rho} \nn \\ 
&~~~~\sum_{\ddot{Q}_{UU'} \in \calQ(Q_{U},Q_{U'})} \mathbb{E}\left[N(\ddot{Q}_{UU'})^{1/\rho}\right]
\exp\{-n E_{2}(\ddot{Q}_{UU'},Q_{S'})/\rho\}.
\end{align}
The expectation in \eqref{REFa3} is upper-bounded as \cite{MERHAV_TYPICAL} 
\begin{align}
&\mathbb{E}\left[N(\ddot{Q}_{UU'})^{1/\rho}\right]  \nn \\
&\lexe \left\{   
\begin{array}{l l}
\exp\{n[2R + R(Q_{S}) + R(Q_{S'}) - I_{\ddot{Q}}(U;U')]/\rho\}   &  \text{  $2R + R(Q_{S}) + R(Q_{S'}) \geq I_{\ddot{Q}}(U;U')$  }\\
\exp\{n[2R + R(Q_{S}) + R(Q_{S'}) - I_{\ddot{Q}}(U;U')] \}   &  \text{  $2R + R(Q_{S}) + R(Q_{S'}) < I_{\ddot{Q}}(U;U')$  } 
\end{array} \right. \\
&=
\exp\{n([2R + R(Q_{S}) + R(Q_{S'}) - I_{\ddot{Q}}(U;U')]_{+}/\rho - [I_{\ddot{Q}}(U;U') - 2R - R(Q_{S}) - R(Q_{S'})]_{+})\},
\end{align}
and so,
\begin{align}
&\left(\mathbb{E}\left[N(\ddot{Q}_{UU'})^{1/\rho}\right] \right)^{\rho} \nn \\
&\lexe
\exp\{n([2R + R(Q_{S}) + R(Q_{S'}) - I_{\ddot{Q}}(U;U')]_{+} - \rho [I_{\ddot{Q}}(U;U') - 2R - R(Q_{S}) - R(Q_{S'})]_{+})\} \\
\label{REFa4}
&\DEF \exp\{n F(\ddot{Q}_{UU'},R,R(Q_{S}),R(Q_{S'}),\rho)\}.
\end{align}
Now, raising \eqref{REFa3} to the power of $\rho$ and substituting \eqref{REFa4} back, we arrive at
\begin{align}
&\left(\mathbb{E} \left[P_{\mbox{\tiny e}}(\calC_{n})^{1/\rho}\right]\right)^{\rho} \nn \\
&\lexe \sum_{Q_{S}} \sum_{Q_{S'}} e^{-n D(Q_{S}\|P_{S})} e^{-n (R + R(Q_{S}))} \nn \\ 
&~~~~\sum_{\ddot{Q}_{UU'} \in \calQ(Q_{U},Q_{U'})} \exp\{n F(\ddot{Q}_{UU'},R,R(Q_{S}),R(Q_{S'}),\rho)\}
\exp\{-n E_{2}(\ddot{Q}_{UU'},Q_{S'})\} \\
&\doteq \exp \left\{-n \cdot \min_{Q_{S},Q_{S'}}  \min_{\ddot{Q}_{UU'} \in \calQ(Q_{U},Q_{U'})} [E_{2}(\ddot{Q}_{UU'},Q_{S'}) - F(\ddot{Q}_{UU'},R,R(Q_{S}),R(Q_{S'}),\rho) \right. \nn \\
&~~~~~~~~~~~~~~~~~\left. + D(Q_{S}\|P_{S}) + R + R(Q_{S})] \right\}.
\end{align}
Hence, it follows from the inequality \cite[Eq.\ (A35)]{TM} 
\begin{align} \label{StartingPoint}
\mathbb{E} [\ln P_{\mbox{\tiny e}}(\calC_{n}) ]
\leq \ln \left( \mathbb{E} \left\{[P_{\mbox{\tiny e}}(\calC_{n})]^{1/\rho} \right\}  \right)^{\rho},
\end{align}
which holds for any $\rho >0$, that
\begin{align}
\liminf_{n \to \infty} -\frac{1}{n} \mathbb{E}[\ln P_{\mbox{\tiny e}}(\calC_{n})] &\geq \min_{Q_{S},Q_{S'}}  \min_{\ddot{Q}_{UU'} \in \calQ(Q_{U},Q_{U'})} [E_{2}(\ddot{Q}_{UU'},Q_{S'}) - F(\ddot{Q}_{UU'},R,R(Q_{S}),R(Q_{S'}),\rho)  \nn \\
&~~~~~~~~~~~~~~~~~ + D(Q_{S}\|P_{S}) + R + R(Q_{S})].
\end{align}
Letting $\rho$ grow without bound yields that
\begin{align}
&\liminf_{n \to \infty} -\frac{1}{n} \mathbb{E}[\ln P_{\mbox{\tiny e}}(\calC_{n})] \nn \\
&\geq \min_{Q_{S},Q_{S'}} \min_{\{\ddot{Q}_{UU'} \in \calQ(Q_{U},Q_{U'}):~I_{\ddot{Q}}(U;U') \leq 2R + R(Q_{S}) + R(Q_{S'})\}} \nn \\
&~~~~~[E_{2}(\ddot{Q}_{UU'},Q_{S'}) + I_{\ddot{Q}}(U;U') - 2R - R(Q_{S}) - R(Q_{S'}) + D(Q_{S}\|P_{S}) + R + R(Q_{S})] \\
&= \min_{Q_{S},Q_{S'}} \min_{\{\ddot{Q}_{UU'} \in \calQ(Q_{U},Q_{U'}):~I_{\ddot{Q}}(U;U') \leq 2R + R(Q_{S}) + R(Q_{S'})\}} \nn \\
&~~~~~[D(Q_{S}\|P_{S}) + E_{2}(\ddot{Q}_{UU'},Q_{S'}) + I_{\ddot{Q}}(U;U') - R - R(Q_{S'})]. 
\end{align}
As a final step, we maximize the exponent function over the design parameters of the code:
\begin{align}
\liminf_{n \to \infty} -\frac{1}{n} \mathbb{E}[\ln P_{\mbox{\tiny e}}(\calC_{n})]
&\geq \max_{\{Q_{UX|S}\}} \min_{Q_{S},Q_{S'}} \min_{\{\ddot{Q}_{UU'} \in \calQ(Q_{U},Q_{U'}):~I_{\ddot{Q}}(U;U') \leq 2R + R(Q_{S}) + R(Q_{S'})\}} \nn \\
&~~~~~[D(Q_{S}\|P_{S}) + E_{2}(\ddot{Q}_{UU'},Q_{S'}) + I_{\ddot{Q}}(U;U') - R - R(Q_{S'})],
\end{align}
which completes the proof of Theorem \ref{THEOREM_TRC}.

\section*{Appendix E - Proof of Theorem \ref{THEOREM_EX}}
\renewcommand{\theequation}{E.\arabic{equation}}
\setcounter{equation}{0}

Assuming that message $m$ was transmitted, the probability of error, for a given code $\calC_{n}$, is given by
\begin{align} \label{DEF_CondError}
P_{\mbox{\tiny e}|m}(\calC_{n})
&= \sum_{Q_{S}} \pr\left[\bS \in \calT(Q_{S})\right] \frac{1}{M(Q_{S})} \sum_{\ell=1}^{M(Q_{S})} \sum_{\bs \in \calT(Q_{S})} P(\bs|\bu_{m,\ell}) \nn \\
&~~\times \sum_{\bx \in \calT(Q_{X|US}|\bu_{m,\ell},\bs)} \frac{1}{|\calT(Q_{X|US}|\bu_{m,\ell},\bs)|} 
\nn \\
&~~\times \sum_{\by \in \calY^{n}} W(\by|\bx,\bs) \IND\left\{\bigcup_{Q_{S'}} \bigcup_{m' \neq m} \bigcup_{\ell'=1}^{M(Q_{S'})} \{G(\hat{P}_{\bu_{m',\ell'}\by},Q_{S'}) \geq G(\hat{P}_{\bu_{m,\ell}\by},Q_{S})\} \right\}. 
\end{align}
Since the whole difference relative to the total probability of error given in \eqref{REF0} is in the averaging over the set of bins (which is of size $M$), we may proceed using the same initial steps as in the proof of Theorem \ref{THEOREM_TRC}, and arrive at (upon using the result in \eqref{REFb2})  
\begin{align}
&P_{\mbox{\tiny e}|m}(\calC_{n}) \nn \\
&\lexe \sum_{Q_{S}} \sum_{Q_{S'}} \pr\left[\bS \in \calT(Q_{S})\right] \frac{1}{M(Q_{S})} \sum_{\ell=1}^{M(Q_{S})}   
\sum_{m' \neq m} 
\sum_{\ell'=1}^{M(Q_{S'})} 
\exp\{-n E_{2}(\hat{P}_{\bu_{m,\ell},\bu_{m',\ell'}},Q_{S'})\} \\
&\doteq \sum_{Q_{S}} \sum_{Q_{S'}} e^{-n D(Q_{S}\|P_{S})} e^{-n R(Q_{S})} \sum_{\ell=1}^{M(Q_{S})}   
\sum_{m' \neq m} 
\sum_{\ell'=1}^{M(Q_{S'})} 
\exp\{-n E_{2}(\hat{P}_{\bu_{m,\ell},\bu_{m',\ell'}},Q_{S'})\} \\
&= \sum_{Q_{S}} \sum_{Q_{S'}} e^{-n D(Q_{S}\|P_{S})} e^{-n R(Q_{S})} 
\sum_{\ddot{Q}_{UU'} \in \calQ(Q_{U},Q_{U'})} \tilde{N}(\ddot{Q}_{UU'})
\exp\{-n E_{2}(\ddot{Q}_{UU'},Q_{S'})\},
\end{align}
where $\calQ(Q_{U},Q_{U'})=\{\ddot{Q}_{UU'}:~\ddot{Q}_{U}=Q_{U},\ddot{Q}_{U'}=Q_{U'}\}$ and
\begin{align}
\tilde{N}(\ddot{Q}_{UU'})
\dfn 
\sum_{\ell=1}^{M(Q_{S})}   
\sum_{m' \neq m} 
\sum_{\ell'=1}^{M(Q_{S'})}
\IND \left\{ (\bu_{m,\ell},\bu_{m',\ell'}) \in \calT(\ddot{Q}_{UU'}) \right\}.
\end{align}
Now, for any $\rho > 1$,
\begin{align}
\mathbb{E} \left[P_{\mbox{\tiny e}|m}(\calC_{n})^{1/\rho}\right]
\label{REFb3}
&\lexe \sum_{Q_{S}} \sum_{Q_{S'}} e^{-n D(Q_{S}\|P_{S})/\rho} e^{-n R(Q_{S})/\rho} \nn \\ 
&~~~~\sum_{\ddot{Q}_{UU'} \in \calQ(Q_{U},Q_{U'})} \mathbb{E}\left[\tilde{N}(\ddot{Q}_{UU'})^{1/\rho}\right]
\exp\{-n E_{2}(\ddot{Q}_{UU'},Q_{S'})/\rho\}.
\end{align}
The expectation in \eqref{REFb3} is upper-bounded as \cite{MERHAV_TYPICAL} 
\begin{align}
&\mathbb{E}\left[\tilde{N}(\ddot{Q}_{UU'})^{1/\rho}\right]  \nn \\
&\lexe \left\{   
\begin{array}{l l}
\exp\{n[R + R(Q_{S}) + R(Q_{S'}) - I_{\ddot{Q}}(U;U')]/\rho\}   &  \text{  $R + R(Q_{S}) + R(Q_{S'}) \geq I_{\ddot{Q}}(U;U')$  }\\
\exp\{n[R + R(Q_{S}) + R(Q_{S'}) - I_{\ddot{Q}}(U;U')] \}   &  \text{  $R + R(Q_{S}) + R(Q_{S'}) < I_{\ddot{Q}}(U;U')$  } 
\end{array} \right. \\
&=
\exp\{n([R + R(Q_{S}) + R(Q_{S'}) - I_{\ddot{Q}}(U;U')]_{+}/\rho - [I_{\ddot{Q}}(U;U') - R - R(Q_{S}) - R(Q_{S'})]_{+})\} \\
\label{REFb4}
&\DEF \exp\{n G(\ddot{Q}_{UU'},R,R(Q_{S}),R(Q_{S'}),\rho)\}.
\end{align}
Substituting \eqref{REFb4} back into \eqref{REFb3} yields a bound which we shall shortly denote by $\Psi(R,\rho)$.

According to Markov's inequality, we get
\begin{align}
\pr \left\{ \frac{1}{M} \sum_{m=1}^{M} P_{\mbox{\tiny e}|m}(\calC_{n})^{1/\rho} > 2 \Psi(R,\rho) \right\} \leq \frac{1}{2},
\end{align}
which means that there exists a code with
\begin{align}
\frac{1}{M} \sum_{m=1}^{M} P_{\mbox{\tiny e}|m}(\calC_{n})^{1/\rho} \leq \Psi(R,\rho).
\end{align}
We conclude that there exists a code $\calC'_{n}$ with $M/2$ bins for which 
\begin{align}
\max_{m} P_{\mbox{\tiny e}|m}(\calC'_{n})^{1/\rho} \leq 4 \Psi(R,\rho),
\end{align}
and so
\begin{align}
&\max_{m} P_{\mbox{\tiny e}|m}(\calC'_{n}) \nn \\
&\lexe \left( \sum_{Q_{S}} \sum_{Q_{S'}} e^{-n D(Q_{S}\|P_{S})/\rho} e^{-n R(Q_{S})/\rho} \right. \nn \\ 
&\left. ~~~~\sum_{\ddot{Q}_{UU'} \in \calQ(Q_{U},Q_{U'})} \exp\{n G(\ddot{Q}_{UU'},R,R(Q_{S}),R(Q_{S'}),\rho)\}
\exp\{-n E_{2}(\ddot{Q}_{UU'},Q_{S'})/\rho\} \right)^{\rho}\\
&\doteq \sum_{Q_{S}} \sum_{Q_{S'}} e^{-n D(Q_{S}\|P_{S})} e^{-n R(Q_{S})} \nn \\ 
& ~~~~\sum_{\ddot{Q}_{UU'} \in \calQ(Q_{U},Q_{U'})} \exp\{n \rho G(\ddot{Q}_{UU'},R,R(Q_{S}),R(Q_{S'}),\rho)\}
\exp\{-n E_{2}(\ddot{Q}_{UU'},Q_{S'})\} \\
&\doteq \exp \left\{-n \cdot \min_{Q_{S},Q_{S'}}  \min_{\ddot{Q}_{UU'} \in \calQ(Q_{U},Q_{U'})} [E_{2}(\ddot{Q}_{UU'},Q_{S'}) - \rho G(\ddot{Q}_{UU'},R,R(Q_{S}),R(Q_{S'}),\rho) \right. \nn \\
&~~~~~~~~~~~~~~~~~\left. + D(Q_{S}\|P_{S}) + R(Q_{S})] \right\},
\end{align}
thus,
\begin{align} \label{ToREF0}
&\liminf_{n \to \infty} - \frac{1}{n} \ln \max_{m} P_{\mbox{\tiny e}|m}(\calC'_{n}) \nn \\
&\geq \min_{Q_{S},Q_{S'}}  \min_{\ddot{Q}_{UU'} \in \calQ(Q_{U},Q_{U'})} [E_{2}(\ddot{Q}_{UU'},Q_{S'}) - \rho G(\ddot{Q}_{UU'},R,R(Q_{S}),R(Q_{S'}),\rho) \nn \\
&~~~~~~~~~~~~~~~~~~~~~~~~~~~~~~~~~~+ D(Q_{S}\|P_{S}) + R(Q_{S})].
\end{align}
Since the inequality in \eqref{ToREF0} holds for every $\rho \geq 1$, the negative exponential rate of the maximal probability of error can be bounded as
\begin{align}
&\liminf_{n \to \infty} - \frac{1}{n} \ln \max_{m} P_{\mbox{\tiny e}|m}(\calC'_{n}) \nn \\
&\geq \sup_{\rho \geq 1} \min_{Q_{S},Q_{S'}}  \min_{\ddot{Q}_{UU'} \in \calQ(Q_{U},Q_{U'})} [E_{2}(\ddot{Q}_{UU'},Q_{S'}) - \rho G(\ddot{Q}_{UU'},R,R(Q_{S}),R(Q_{S'}),\rho) \nn \\
&~~~~~~~~~~~~~~~~~~~~~~~~~~~~~~~~~~+ D(Q_{S}\|P_{S}) + R(Q_{S})] \\
&\geq \lim_{\rho \to \infty} \min_{Q_{S},Q_{S'}}  \min_{\ddot{Q}_{UU'} \in \calQ(Q_{U},Q_{U'})} [E_{2}(\ddot{Q}_{UU'},Q_{S'}) - \rho G(\ddot{Q}_{UU'},R,R(Q_{S}),R(Q_{S'}),\rho) \nn \\
&~~~~~~~~~~~~~~~~~~~~~~~~~~~~~~~~~~+ D(Q_{S}\|P_{S}) + R(Q_{S})].
\end{align}
At this point, we shall proceed along the same steps as in \cite[Eqs.\ (A.26)-(A.37)]{TM_UD}, and conclude that
\begin{align}
&\liminf_{n \to \infty} - \frac{1}{n} \ln \max_{m} P_{\mbox{\tiny e}|m}(\calC'_{n}) \nn \\
&\geq \min_{Q_{S},Q_{S'}} \min_{\{\ddot{Q}_{UU'} \in \calQ(Q_{U},Q_{U'}):~I_{\ddot{Q}}(U;U') \leq R + R(Q_{S}) + R(Q_{S'})\}} \nn \\
&~~~~~[E_{2}(\ddot{Q}_{UU'},Q_{S'}) + I_{\ddot{Q}}(U;U') - R - R(Q_{S}) - R(Q_{S'}) + D(Q_{S}\|P_{S}) + R(Q_{S})] \\
&= \min_{Q_{S},Q_{S'}} \min_{\{\ddot{Q}_{UU'} \in \calQ(Q_{U},Q_{U'}):~I_{\ddot{Q}}(U;U') \leq R + R(Q_{S}) + R(Q_{S'})\}} \nn \\
&~~~~~[D(Q_{S}\|P_{S}) + E_{2}(\ddot{Q}_{UU'},Q_{S'}) + I_{\ddot{Q}}(U;U') - R - R(Q_{S'})]. 
\end{align}
As a final step, we maximize the exponent function over the design parameters of the code:
\begin{align}
\liminf_{n \to \infty} - \frac{1}{n} \ln \max_{m} P_{\mbox{\tiny e}|m}(\calC'_{n})
&\geq \max_{\{Q_{UX|S}\}} \min_{Q_{S},Q_{S'}} \min_{\{\ddot{Q}_{UU'} \in \calQ(Q_{U},Q_{U'}):~I_{\ddot{Q}}(U;U') \leq R + R(Q_{S}) + R(Q_{S'})\}} \nn \\
&~~~~~[D(Q_{S}\|P_{S}) + E_{2}(\ddot{Q}_{UU'},Q_{S'}) + I_{\ddot{Q}}(U;U') - R - R(Q_{S'})],
\end{align}
which completes the proof of Theorem \ref{THEOREM_EX}.


\begin{thebibliography}{AA}
	
	\bibitem{ArikanMerhav1998}
	E.~Arikan and N.~Merhav, ``Guessing subject to distortion,'' {\it IEEE Trans.\ Inf.\ Theory}, vol.\ 44, no.\ 3, pp.\ 1041--1056, May 1998.
	
	\bibitem{AM2018}
	R.~Averbuch and N.~Merhav, ``Exact random coding exponents and universal decoders for the asymmetric broadcast channel,'' {\it IEEE Trans.\ Inf.\ Theory}, vol.\ 64, no.\ 7, pp.\ 5070--5086, July 2018.
	
	\bibitem{BargForney}
	A.~Barg and G.~D.~Forney, Jr., ``Random codes: minimum distances and error exponents,'' {\it IEEE Trans. Inf. Theory}, vol. 48, no. 9, pp. 2568--2573, September 2002.
	
	\bibitem{Duality2}
	R.~J.~Barron, B.~Chen, and G.~W.~Wornell, ``The duality between information embedding and source coding with side information and some applications,'' {\it IEEE Trans. Inf. Theory}, vol. 49, no. 5, pp. 1159--1180, May 2003.
	
	\bibitem{CGF2022}
	G.~Cocco, A.~Guill\'en i F\`abregas, and J.~Font-Segura,
	``Typical error exponents:
	a dual domain derivation,'' 2022. [Online]. Available: {\tt https://arxiv.org/abs/2203.15607}
	
	\bibitem{Costa1983}
	M.~Costa, ``Writing on dirty paper,'' {\it IEEE Trans. Inf. Theory}, vol. 29, no. 3, pp. 439--441, May 1983.
	
	\bibitem{Duality1}
	T.~M.~Cover and M.~Chiang, ``Duality between channel capacity and rate distortion with side information,'' {\it IEEE Trans. Inf. Theory}, vol. 48, no. 6, pp. 1629--1638, June 2002.
	
	\bibitem{El-Gamal}
	A. El Gamal and Y. H. Kim, {\it Network Information Theory.} Cambridge, U.K.: Cambridge Univ. Press, 2011.
	
	\bibitem{ErezZamir2001}
	U.~Erez and R.~Zamir, ``Error exponents of modulo-additive noise channels with side information at the transmitter,'' {\it IEEE Trans. Inf. Theory}, vol. 47, no. 1, pp. 210--218, January 2001.
	
	\bibitem{GAL68}
	R.~G.~Gallager, {\it Information Theory and Reliable Communication}, New York, Wiley 1968.
	
	\bibitem{GP1980}
	S.~I.~Gel'fand and M.~S.~Pinsker, ``Coding for channel with random parameters," {\it Problem of Control and Information Theory}, vol. 9, no. 1, pp. 19--31, 1980.
	
	\bibitem{HEG1983}
	C.~Heegard and A.~El Gamal, ``On the capacity of computer memory with defects,,'' {\it IEEE Trans. Inf. Theory}, vol. 29, no. 5, pp. 731--739, September 1983.
	
	\bibitem{KSM2007}
	G.~Keshet, Y.~Steinberg, and N.~Merhav, ``Channel coding in the presence of side information,'' {\it Foundations and Trends in Communications and Information Theory,} vol. 4, nos. 6, pp. 1--136, 2007.
	
	\bibitem{LMK2006}
	T.~Liu, P.~Moulin, and R.~Koetter, ``On error exponents of modulo lattice additive noise channels," {\it IEEE Trans. Inf. Theory}, vol. 52, no. 2, pp. 454--471, February 2006.
	
	\bibitem{LW2006}
	T.~Liu and P.~Viswanath, ``Opportunistic orthogonal writing on dirty paper," {\it IEEE Trans. Inf. Theory}, vol. 52, no. 5, pp. 1828--1846, May 2006.	
	
	\bibitem{MERHAV1993}
	N.~Merhav, ``Universal decoding for memoryless Gaussian channels with a deterministic interference," {\it IEEE Trans. Inf. Theory}, vol. 39, no. 4, pp. 1261--1269, July 1993.
	
	\bibitem{MERHAV09}
	N.~Merhav, ``Statistical physics and information theory,'' {\it Foundations and Trends in Communications and Information Theory,} vol. 6, nos. 1-2, pp. 1--212, 2009.
	
	\bibitem{MERHAV2013}
	N.~Merhav, ``Universal decoding for arbitrary channels relative to a given class of decoding metrics," {\it IEEE Trans. Inf. Theory}, vol. 59, no. 9, pp. 5566--5576, September 2013.
	
	\bibitem{MERHAV2014}
	N.~Merhav, ``Exact random coding error exponents of optimal bin index decoding," {\it IEEE Trans. Inf. Theory}, vol. 60, no. 10, pp. 6024--6031, October 2014.
	
	\bibitem{MERHAV2017}
	N.~Merhav, ``Reliability of universal decoding based on vector-quantized codewords," {\it IEEE Trans. Inf. Theory}, vol. 63, no. 5, pp. 2696--2709, May 2017.
	
	\bibitem{MERHAV_TYPICAL}
	N.~Merhav, ``Error exponents of typical random codes," {\it IEEE Trans. Inf. Theory}, vol. 64, no. 9, pp. 6223--6235, September 2018.
	
	\bibitem{MERHAV_GAUSS}
	N. Merhav, ``Error exponents of typical random codes for the colored Gaussian channel," {\it IEEE Trans. Inf. Theory}, vol. 65, no. 12, pp. 8164--8179, December 2019.
	
	\bibitem{MERHAV_TRELLIS}
	N. Merhav, ``Error exponents of typical random trellis codes," {\it IEEE Trans. Inf. Theory}, vol. 66, no. 4, pp. 2067--2077, April 2020.
	
	\bibitem{MERHAV_IID}
	N. Merhav, ``A Lagrange--dual lower bound to the error exponent of the typical random code," {\it IEEE Trans. Inf. Theory}, vol. 66, no. 6, pp. 3456--3464, June 2020.
	
	\bibitem{MERHAV_Shamai_2003}
	N.~Merhav and S.~Shamai, ``On joint source-channel coding for the Wyner-Ziv source and the Gel'fand-Pinsker channel,," {\it IEEE Trans. Inf. Theory}, vol. 49, no. 11, pp. 2844--2855, November 2003.
	
	\bibitem{MoulinOsullivan2003}
	P.~Moulin and J.~A.~O'Sullivan, ``Information-theoretic analysis of information
	hiding," {\it IEEE Trans. Inf. Theory}, vol. 49, no. 3, pp. 563--593, March 2003.
	
	\bibitem{Moulin2007}
	P.~Moulin and Y.~Wang, ``Capacity and random-coding exponents for channel coding with side information," {\it IEEE Trans. Inf. Theory}, vol. 53, no. 4, pp. 1326--1347, April 2007.
	
	\bibitem{PRAD2014}
	A.~Nazari, A.~Anastasopoulos, and S.~S.~Pradhan, ``Error exponent for multiple--access channels: lower bounds," {\it IEEE Trans. Inf. Theory}, vol. 60, no. 9, pp. 5095--5115, September 2014.
	
	\bibitem{Duality3}
	S.~S.~Pradhan and J.~C.~Ramchandran, ``Duality between source coding and channel coding and its extension to the side information case,'' {\it IEEE Trans. Inf. Theory}, vol. 49, no. 5, pp. 1181--1203, May 2003.
	
	\bibitem{SM2004}
	A.~Somekh--Baruch and N.~Merhav, ``On the random coding error exponents of the single-user and the multiple-access Gel'fand--Pinsker channels,'' in {\it Proc. IEEE Int. Symp. Information Theory}, Chicago, IL, June/July 2004, p. 448.
	
	\bibitem{Multiaccess}
	A.~Somekh--Baruch, S.~Shamai, and S.~Verd\'u, ``Cooperative multiple-access encoding with states
	available at one transmitter,'' {\it IEEE Trans. Inf. Theory}, vol. 54, no. 10, pp. 4448--4469, October 2008.
	
	\bibitem{Broadcast}
	Y.~Steinberg, ``Coding for the degraded broadcast channel with random parameters, with causal and noncausal side information,'' {\it IEEE Trans. Inf. Theory}, vol. 51, no. 8, pp. 2867--2877, August 2005.

	\bibitem{TM}
	R.~Tamir and N.~Merhav, ``Trade-offs between error exponents and excess--rate exponents of typical Slepian--Wolf codes,'' {\it Entropy} 2021, 23, 265. {\tt http://doi.org/10.3390/e23030265}.

	\bibitem{TM_UD}
	R.~Tamir and N.~Merhav, ``Universal decoding for the typical random code and for the expurgated code,'' {\it IEEE Trans.\ Inf.\ Theory}, vol.\ 68, no.\ 4, pp.\ 2156--2168, April 2022.
	
	\bibitem{TMWG}
	R.~Tamir, N.~Merhav, N.~Weinberger, and A.~Guill\'en i F\`abregas,
	``Large deviations behavior of the logarithmic error probability of random codes,'' {\it IEEE Trans.\ Inf.\ Theory}, vol.\ 66, no.\ 11, pp.\ 6635--6659, November 2020.
	
	\bibitem{TCFG2022}
	L.~V.~Truong, G.~Cocco, J.~Font-Segura, and A.~Guill\'en i F\`abregas,
	``Concentration properties of random codes,'' 2022. [Online]. Available: {\tt https://arxiv.org/abs/2203.07853}
	
	\bibitem{TN2009}
	H.~Tyagi and P.~Narayan, ``The Gelfand-Pinsker channel: strong converse and upper bound for the reliability function,'' in {\it Proc. IEEE Int. Symp. Information Theory}, Seoul, Korea, June/July 2009, pp. 1954--1957.
	
	\bibitem{WYNER}
	A.~D.~Wyner, ``A bound on the number of distinguishable functions which
	are time-limited and approximately band-limited,''
	{\it SIAM Journal on Applied Mathematics},
	vol.~24, no.~3, pp.~289--297, May 1973.
	
\end{thebibliography}
\end{document}